\documentclass[preprint,11pt]{elsarticle}
\usepackage{epstopdf}
\usepackage{amsmath}
\usepackage{amssymb}
\usepackage{amsbsy}
\usepackage{makeidx}
\usepackage{float}
\usepackage{graphicx}
\usepackage{amsfonts}
\usepackage{placeins}
\usepackage[caption = false]{subfig}
\usepackage{wasysym}
\usepackage{comment}
\usepackage{color}
\biboptions{numbers,sort&compress}
\def\sech{\mathop{\rm sech}\nolimits}

\graphicspath{{figures/}}

\begin{document}

\begin{frontmatter}

\title{
Solitons in Multi-Component Nonlinear Schr{\"o}dinger Models: A Survey of Recent Developments}

\author[amherst]{P.~G. Kevrekidis} 
\author[athens]{D.~J. Frantzeskakis}

\address[amherst]{Department of Mathematics and Statistics,
University of Massachusetts, Amherst MA 01003-4515, USA}
\address[athens]{Department of Physics, National and Kapodistrian University of Athens, Panepistimiopolis, Zografos, Athens 15784, Greece}

\begin{abstract}
In this review we try to capture some of the recent excitement induced
by experimental developments, but also by a large volume of
theoretical and computational studies addressing multi-component nonlinear
Schr{\"o}dinger models and the localized structures that they support.
We focus on some prototypical structures, namely
the dark-bright and dark-dark solitons.
Although our focus will be on one-dimensional, two-component
Hamiltonian models, we also discuss 
variants, including
three (or more)-component models, higher-dimensional 
states, as well as dissipative settings.
We also offer an outlook on 
interesting possibilities for future work on this theme.
\end{abstract}

\end{frontmatter}

\section{Introduction}

Since the early days of nonlinear science and the explosion of interest in integrable models,
it was realized that the nonlinear Schr{\"o}dinger (NLS) equation~\cite{sulem,ablowitz,siambook}
is a universal model describing envelope solitons in dispersive nonlinear media; as such,
it plays a central role in a variety of contexts, ranging from water waves 
and plasmas~\cite{infeld} to nonlinear optics~\cite{kivshar} and atomic Bose-Einstein
condensates (BECs)~\cite{emergent}.
Furthermore, already from the 70s, it was found that
the interaction of waves of different frequencies
gives rise to vector (multi-component) NLS models~\cite{manakov,intman1}.
It is thus not surprising that since then a considerable volume
of work was dedicated into multi-component NLS equation settings.
Arguably, the workhorse of many
relevant studies was the Manakov system~\cite{manakov} (which is integrable~\cite{intman2}), characterized by equal nonlinear
interactions within and between components.

Vector solitons of this model have attracted much attention, especially in
the setting of defocusing intra- and inter-component interactions. In such a
case, of particular interest are dark-bright (DB) solitons. In these structures,
the bright soliton, which would not exist in the defocusing setting,
only emerges because of an effective potential well created by the dark soliton
through the inter-component interaction; as such, DB solitons can be thought of
as ``symbiotic'' structures. DB solitons have 
attracted much attention~\cite{christo,vdbysk1,vddyuri,ralak,dbysk2,shepkiv,parkshin},
especially due to potential applications in optics, where dark solitons
could be used as adjustable waveguides for weak signals
(see, e.g., Ref.~\cite{kivshar} and references therein).
Importantly, these early theoretical developments, focusing on
the integrable theory, its exact solutions and perturbations 
thereof, were also accompanied by pioneering experiments
in photorefractive 
media, where DB solitons
were observed and studied~\cite{seg1,seg2}.

Here, we target this specific multi-component setting,
featuring defocusing inter- and intra-component nonlinearities,
and the corresponding solitary waves
(below, we use the term ``soliton'' in a loose sense,
without implying complete integrability);
for focusing multi-component systems and
bright solitons see, e.g., Refs.~\cite{placeholder,emergent}
and references therein.
We will also consider atomic BECs~\cite{stringari,roman}, which
have provided a new spark for this theme~\cite{siambook,emergent}.
Indeed, seminal experiments have realized
multi-component BECs,
as mixtures of, e.g.,
different spin states of the same atom species
({\it pseudo-spinor} condensates)~\cite{Hall1998a,chap01:stamp}, or
different Zeeman sub-levels of the same hyperfine level 
({\it spinor} condensates)~\cite{Stenger1998a,kawueda,stampueda}.
In BECs, the soliton
in one species can be the same or different
to that in the other species. 
Of particular interest here 
will be vector solitons where one component is a dark soliton.

Our presentation is structured as follows. In Sec.~2, we present the model and
discuss its experimental motivation. In Sec.~3, we analyze statics and dynamics
of single and multiple vector solitons. In Sec.~4, we discuss various settings
and parameter regimes for vector solitons. Finally, in Sec.~5, we briefly summarize
our conclusions and discuss future challenges.

\section{Background and experimental motivation}

\subsection{The multi-component NLS model}

A mixture of $\mathcal N$ bosonic components can be described, at the
mean-field level~\cite{stringari}, by 
a system of $\mathcal N$ coupled NLS equations
[alias Gross-Pitaevskii equations (GPEs) in this context].
When the different components pertain
to the same atom species, this system reads
(see, e.g., Refs.~\cite{emergent,siambook}):
\begin{equation}
i \frac{\partial \psi_n}{\partial t}=-\frac{1}{2} \nabla^2 \psi_n
+ V_n({\bf r}) \psi_n + \sum_{k=1}^{\mathcal N}\left[ g_{nk}|\psi_k|^2 \psi_n
-\kappa_{nk} \psi_k + (\Delta \mu_{nk}) \psi_n  \right].
\label{2C}
\end{equation}
Here, $\psi_n$ denotes the wavefunction of the $n$-th component
($n=1,\dots,{\mathcal N}$),
$V_n({\bf r})$ 
is the trapping potential confining the $n$-th component
which is typically parabolic,
$(\Delta \mu_{nk})$ is the 
chemical potential (eigenvalue parameter) difference
between components $n$ and $k$,
the nonlinearity coefficients $g_{nk}=g_{kn}$ characterize inter-atomic collisions, while
the linear coupling coefficients $\kappa_{nk}=\kappa_{kn}$ are responsible for
spin state inter-conversion, induced typically by a
spin-flipping resonant electromagnetic wave~\cite{NZ}.
This system conserves the energy $E$ and the total number of atoms,
$N\equiv \sum_{k=1}^{\mathcal N} N_{k}=\sum_{k=1}^{\mathcal N}\int |\psi_k|^2 d\mathbf{r}$;
furthermore, in the absence of linear inter-conversions (i.e., $\kappa_{nk}=0$),
the number of atoms of each component $N_{k}$ is conserved.

The principal paradigm on which our exposition will be based
is that of two bosonic species (${\mathcal N}=2$), where we will
assume that the system is homogeneous ($V_n=0$) or trapped ($V_n \neq 0$).
In the homogeneous case with $\kappa_{nk}=0$ and $(\Delta\mu_{nk})=0$,
the binary mixture is {\it immiscible}
provided that the following {\it immiscibility condition} holds~\cite{miscibility}:
\begin{equation}
\Delta \equiv (g_{12}^2-g_{11}g_{22})/g_{11}^{2} > 0,
\label{Delta}
\end{equation}
where $\Delta$ is the so-called miscibility parameter.
In the experiments,
this parameter 
assumes values of the order
of $10^{-3}$ or even less; for example, $\Delta \approx 9 \times 10^{-4}$ or $\Delta \approx 0.036$
for a mixture of two spin states of $^{87}$Rb~\cite{Hall1998a}
or $^{23}$Na~\cite{chap01:stamp} BEC, respectively.
Condition~(\ref{Delta}) 
corresponds to the
case where
the mutual repulsion between species is stronger than the repulsion
between atoms of the same species. 
Then, the two species do not mix and instead
tend to separate by filling two different spatial regions, thus forming,
e.g., a ``ball and shell'' configuration (cf.~experiment of Ref.~\cite{Hall1998a}),
or two 
domain-wall structures of a similar type, one in
each component~\cite{Marek}.

In what follows, we will chiefly operate in the
vicinity of this threshold which favors the emergence of
DB solitons, and use $\kappa_{nk}=0$. This setting is relevant both to experiments and
to the mathematically tractable Manakov limit of $g_{nk}=1$~\cite{manakov}.
Experimental results have demonstrated that the
prototypical vector
solitons that may be supported in such quasi one-dimensional (1D) systems
involve 
a dark soliton in one component, while the second one may be either
a bright soliton, so that the vector soliton is a dark-bright (DB) soliton~\cite{hamburg,pe1,pe2,pe3,azu},
or a dark soliton,
so that the vector soliton is a dark-dark (DD) soliton~\cite{pe4,pe5}.
It should also be noted that, in theory, dark-antidark solitons (the latter being
humps, instead of dips, 
on top of the
background state), 
have also been predicted~\cite{EPJD,hadi,jap}.

\subsection{Optics experiments}

\begin{figure}
	\begin{center}
\includegraphics[height=3.1cm]{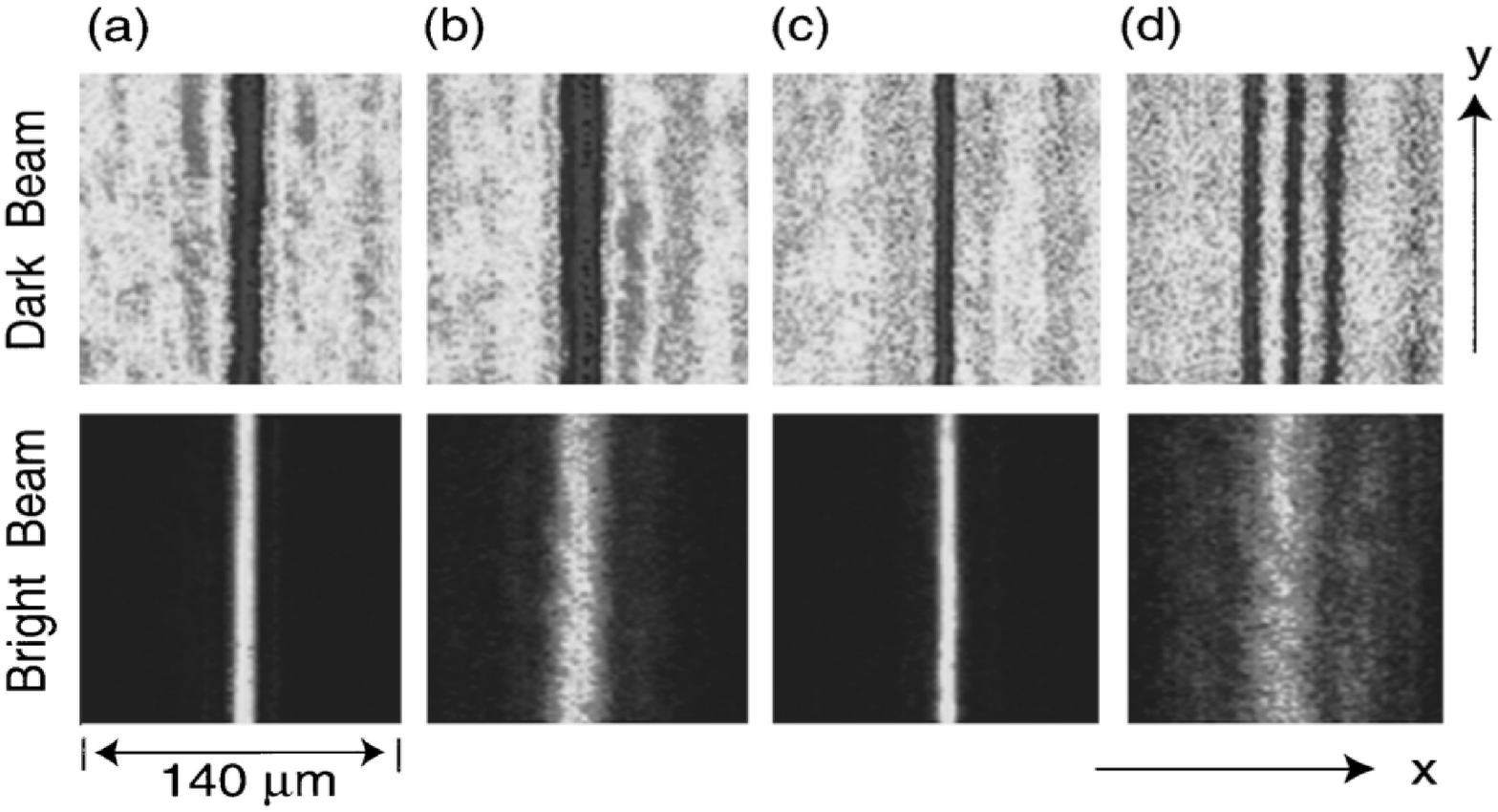}
~
\includegraphics[height=3.1cm]{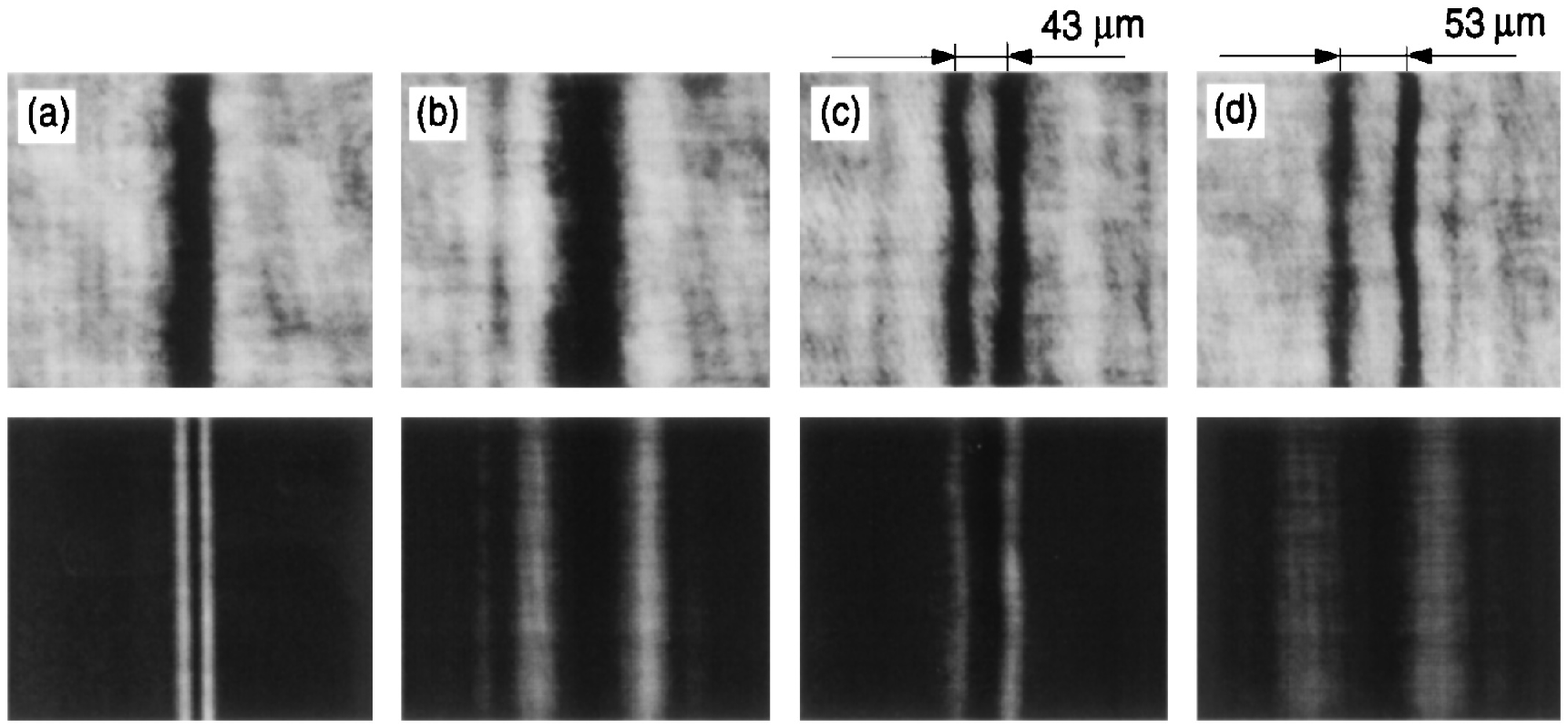}
\caption{(Color Online)
Formation of DB solitons in photorefractive crystals:
the left four panels (adapted from Ref.~\cite{seg1}) showcase
the evolution of an initial condition [panel (a)],
upon propagation under linear evolution, leading to
dispersion [panel (b)], nonlinear evolution of uncoupled components, again
leading to breakup/dispersion [panel (d)], and under coupled nonlinear
evolution [panel (c)].
A similar case example, but for two bright beams,
is shown in the right four panels (adapted from Ref.~\cite{seg2}).
The dark (bright) component is shown in the top (bottom) panel.
}
		\label{revip_fig1}
	\end{center}
\end{figure}

As discussed above, DB soliton states were first observed in pioneering experiments
in optics~\cite{seg1,seg2};
their key findings, summarized also in Fig.~\ref{revip_fig1},
were as follows. Experiments were performed in a photorefractive
(strontium barium niobate) medium, with an input corresponding to a dark soliton (created through
an optical mask) in one component, coupled to a bright soliton
in the second component 
[panel (a) in the left of the figure].
For low intensity, i.e., for a linear evolution, both components
underwent dispersion-induced broadening [panel (b)]. On the other hand,
uncoupled nonlinear evolution 
[panel (d)] again resulted in 
dispersion of the bright component, 
since the corresponding structure 
was not effectively confined by the presence of the dark one.
Only when the evolution is nonlinear {\it and} the two components are
coupled 
[panel (c)], is it possible for the DB soliton state to persist.
Similarly, for initial conditions conducive to a breakup
into two DB waves (right four panels in Fig.~\ref{revip_fig1}),
linear propagation, and 
uncoupled nonlinear propagation 
[panels (b) and (d)] do not lead to coherent states. Only the
coupled nonlinear evolution [panel (c)] leads to a robust
pair of DB solitons, i.e., a so-called ``solitonic gluon''~\cite{seg2}, or ``soliton molecule''.

\subsection{BEC experiments}

\begin{figure}
	\begin{center}
\includegraphics[height=6cm]{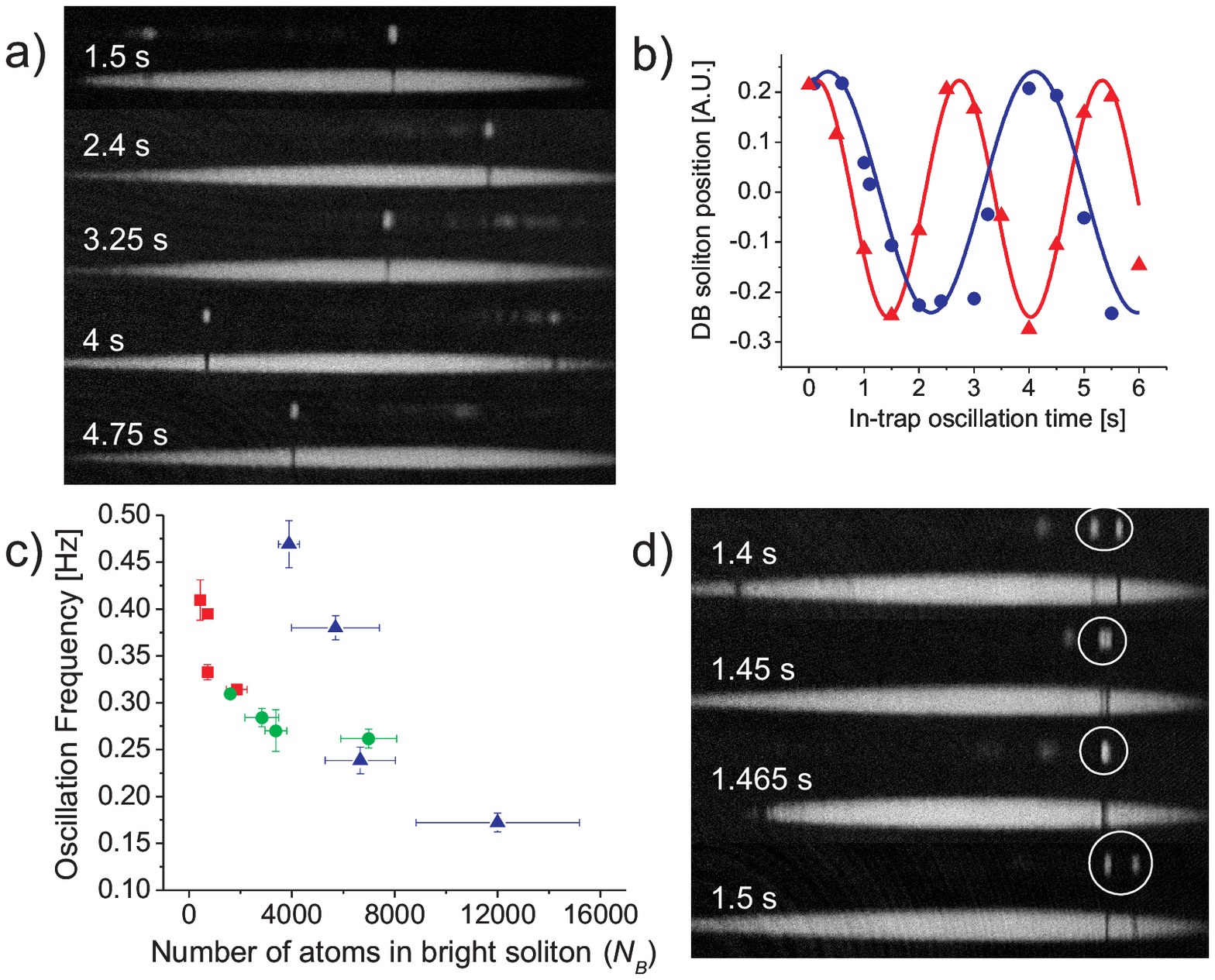}
~
\includegraphics[height=6cm]{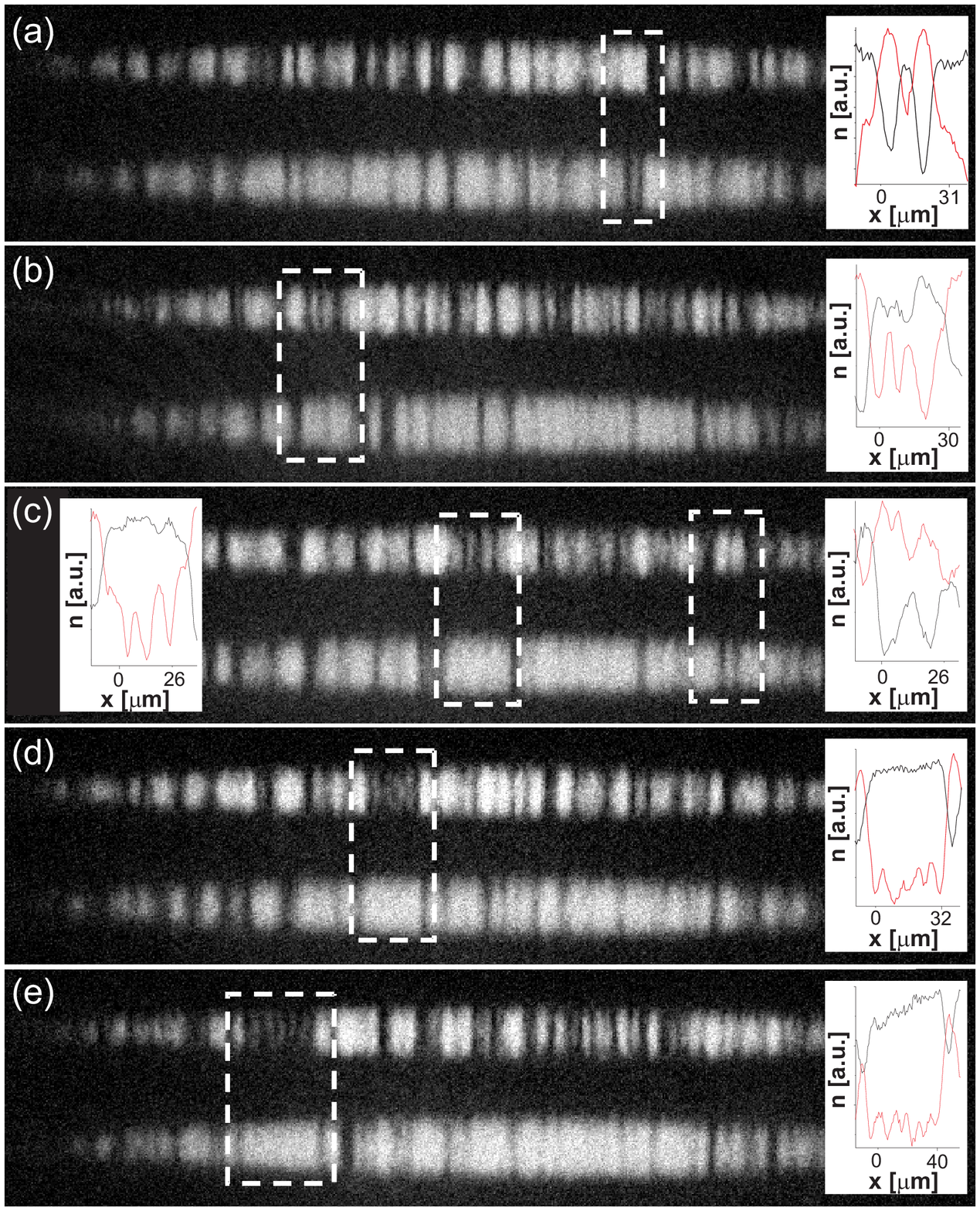}
\caption{(Color Online) The left four panels
(adapted from Ref.~\cite{pe2}),
show: the oscillation of a DB soliton in
a trapped BEC (a); the soliton position 
for different ``masses'' of the bright soliton (b); the oscillation
frequency as a function of the bright soliton mass (c) and the
collision of two DB solitons (d). The right panel
(adapted from Ref.~\cite{pe3}) shows 
(transient) states bearing 2 (a), 3 (b), 4 (c), 5 (d) and even 6 (e) DB
solitons.
}
		\label{revip_fig2}
	\end{center}
\end{figure}

In the context of BECs, DB solitons 
were first predicted and studied in binary BECs in Ref.~\cite{buschanglin}.
While this theoretical work triggered a number of follow-up studies 
ramifying the original idea~\cite{siambook}, 
arguably, it was the Hamburg experiment~\cite{hamburg}, and subsequently those
at Pullman~\cite{pe1,pe4,pe2,pe3,pe5,azu} that put the topic in a fundamentally new
perspective by revealing the experimental possibilities thereof.
In particular, in Ref.~\cite{hamburg} it was demonstrated that DB solitons
can be created by a phase-imprinting technique 
in the two-component setting, and 
also showcased
their robust oscillations in a quasi-1D parabolic trap. In turn,
a different breed of experiments was introduced later, where both DB~\cite{pe1,pe2,pe3}
and DD~\cite{pe4,pe5} solitons
were generated spontaneously via instability mechanisms
in 
counterflow experiments: specifically,
the condensates (composed of two distinct hyperfine states) were
spatially separated and subsequently ``slammed'' against
each other, leading to the spontaneous
emergence of the coherent structures of interest.

\begin{figure}
	\begin{center}
\includegraphics[height=4.4cm]{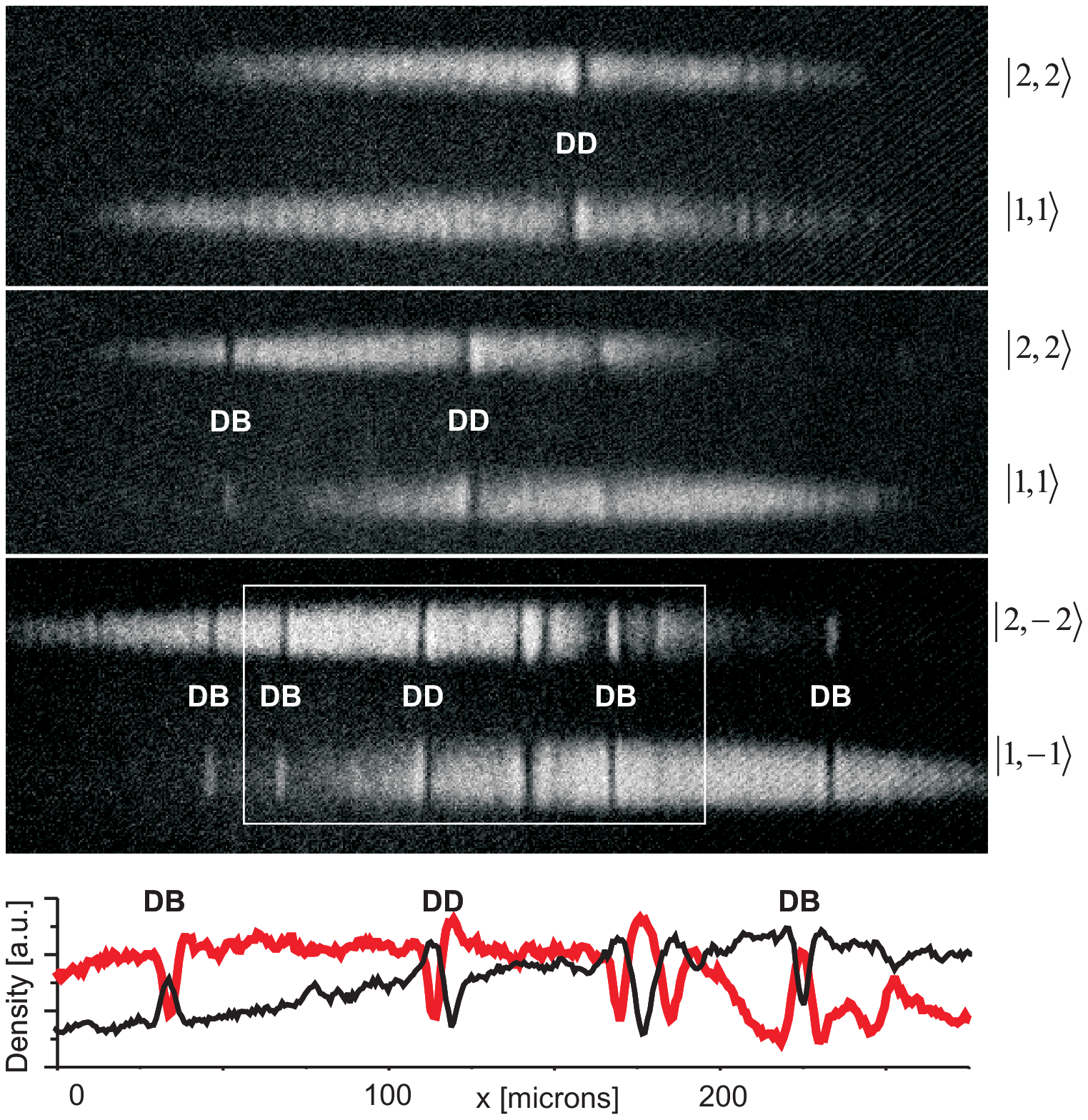}
\includegraphics[height=4.4cm]{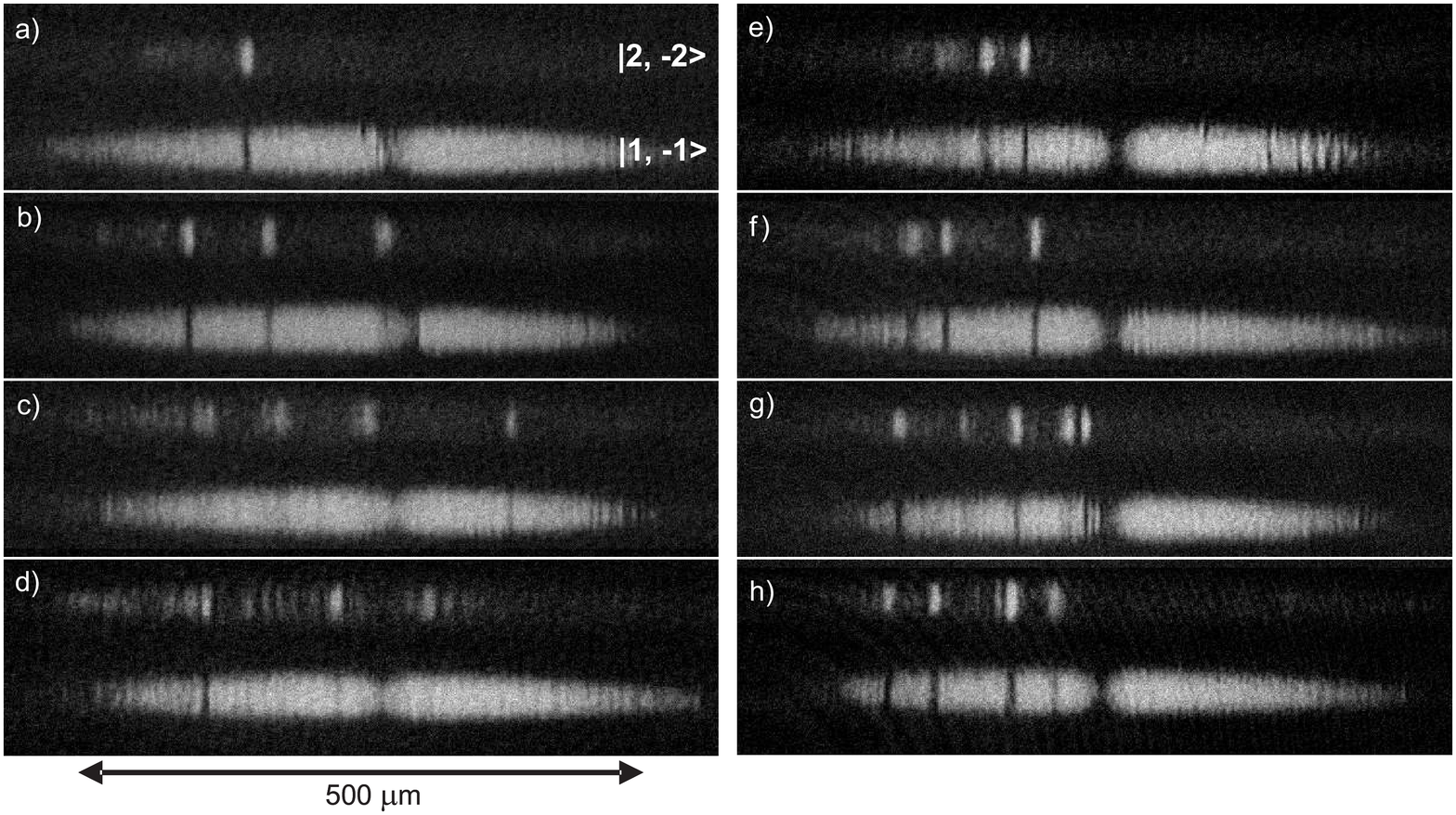}
\caption{(Color Online) The left panels (adapted from Ref.~\cite{pe5})
depict different experimental snapshots of
counterflow experiments of two different BEC components; 
the density profiles shown in
the fourth panel correspond to 
a spatial region depicted in the third panel. The eight panels on the right
(adapted from Ref.~\cite{azu}) show progressively in time
the interaction of a DB soliton 
with a Gaussian barrier: for a sufficiently high energy the wave is transmitted [panels (a-d)],
while for a lower energy it is reflected [panels (e-h)].
}
		\label{revip_fig3}
	\end{center}
\end{figure}

Examples illustrating the above mentioned experimental
possibilities and findings are
shown in Figs.~\ref{revip_fig2} and~\ref{revip_fig3}.
In particular, Fig.~\ref{revip_fig2} 
depicts robust oscillations of a DB soliton 
in a quasi-1D trap [left panel (a)].
The time-series of the wave center position [left panel (b)]
allows to infer the oscillation frequency, 
as well as its dependence on the ``mass'' (number of atoms) of the bright component
[left panel (c)]; this suggests that 
heavier solitons 
become slower, bearing an increasing period.
In addition, collisions between two
separately oscillating DB solitons
are shown [left panel (d)]. The right set of panels
depicts that states with two up to six DB
solitons can (at least transiently)
form during the complex evolution of the counterflow experiments.
This again points to the idea of soliton molecules and progressively
more elaborate states bearing multiple DB solitons. 

Finally, Fig.~\ref{revip_fig3} motivates some variants on the theme of DB
solitons in BECs, by virtue of additional experiments~\cite{pe3,azu}.
The left panel shows that in the counterflow experiments not only DB
solitons but also DD ones are generated.
The middle and right panels depict the
interaction of a DB soliton 
with a potential barrier, 
induced by a Gaussian laser beam.
When the barrier is shallow (or equivalently 
the wave bears sufficient energy to overcome the barrier), transmission
through the barrier is observed (middle panel), while 
in the reverse scenario of a deep barrier (or insufficiently energetic DB solitons), 
near perfect reflection is realized (right panel). 

\section{Dark-Bright and Dark-Dark Solitons in 1D
}

\subsection{The homogeneous setting}

Motivated by these experimental results, we now turn to a theoretical study
of the DB and DD solitons 
in the 1D
setting. There, in the case of two components and
in the absence of linear coupling, Eq.~(\ref{2C})
becomes:
\begin{equation}
\begin{array}{rcl}
i \partial_t \psi_1  &=& -\frac{1}{2}\, \partial_{x}^2 \psi_1 + V(x)\psi_1
+(|\psi_1|^2 + |\psi_2|^2 -\mu_1) \psi_1,
\\[1.5ex]
i\partial_t \psi_2  &=& -\frac{1}{2}\, \partial_{x}^2 \psi_2 +V(x)\psi_2
+ (|\psi_1|^2 + |\psi_2|^2- \mu_2) \psi_2,
\end{array}
\label{deq12}
\end{equation}
where we have considered the Manakov limit with 
$g_{ij}=1$~\cite{manakov}. 
This is motivated, e.g., by the physically relevant case of the
$^{87}$Rb BEC, where hyperfine states are characterized by
almost equal coupling strengths~\cite{hamburg,pe1}.
For $\mu_1=\mu_2=\mu$, and neglecting ---to a first approximation---
the confining potential, Eqs.~(\ref{deq12}) possess the following
DB soliton solution:
\begin{eqnarray}
\psi_1(x,t) &=& \sqrt{\mu} (\cos\phi\,\tanh \xi+i \sin\phi),
\label{dark_part}
\\
\psi_2(x,t) &=& \eta\, \sech \xi\, \exp[ikx+i\theta(t)],
\label{bright_part}
\end{eqnarray}
where $\xi=D(x-x_0(t))$, $\phi$ is the phase angle of the dark soliton,
$\cos\phi$ and $\eta$ are the amplitudes of the dark and bright
solitons, while $D$ and $x_0(t)$ describe the inverse width
and the center position of the DB soliton. For this solution to be valid,
$D^2=\mu \cos^2 \phi - \eta^2$, the soliton velocity
is given by $\dot{x}_0=k=D\tan\phi$,
and phase $\theta(t)=(1/2) (D^2-k^2) t + \theta_0$ (with $\theta_0={\rm const.}$).

\begin{figure}
	\begin{center}
		\begin{tabular}{cc}
			\includegraphics[scale=0.2]{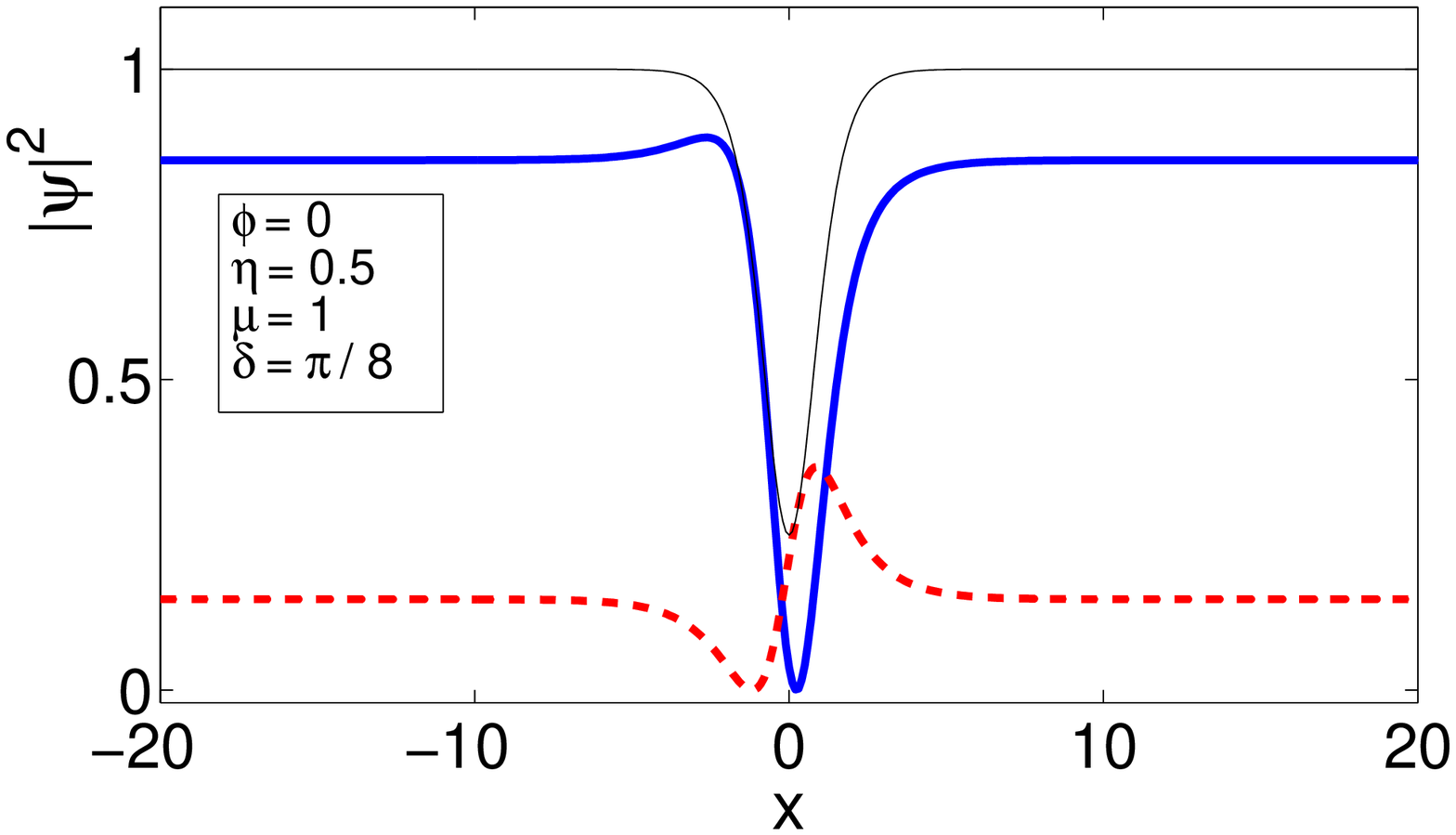}
			\includegraphics[scale=0.2]{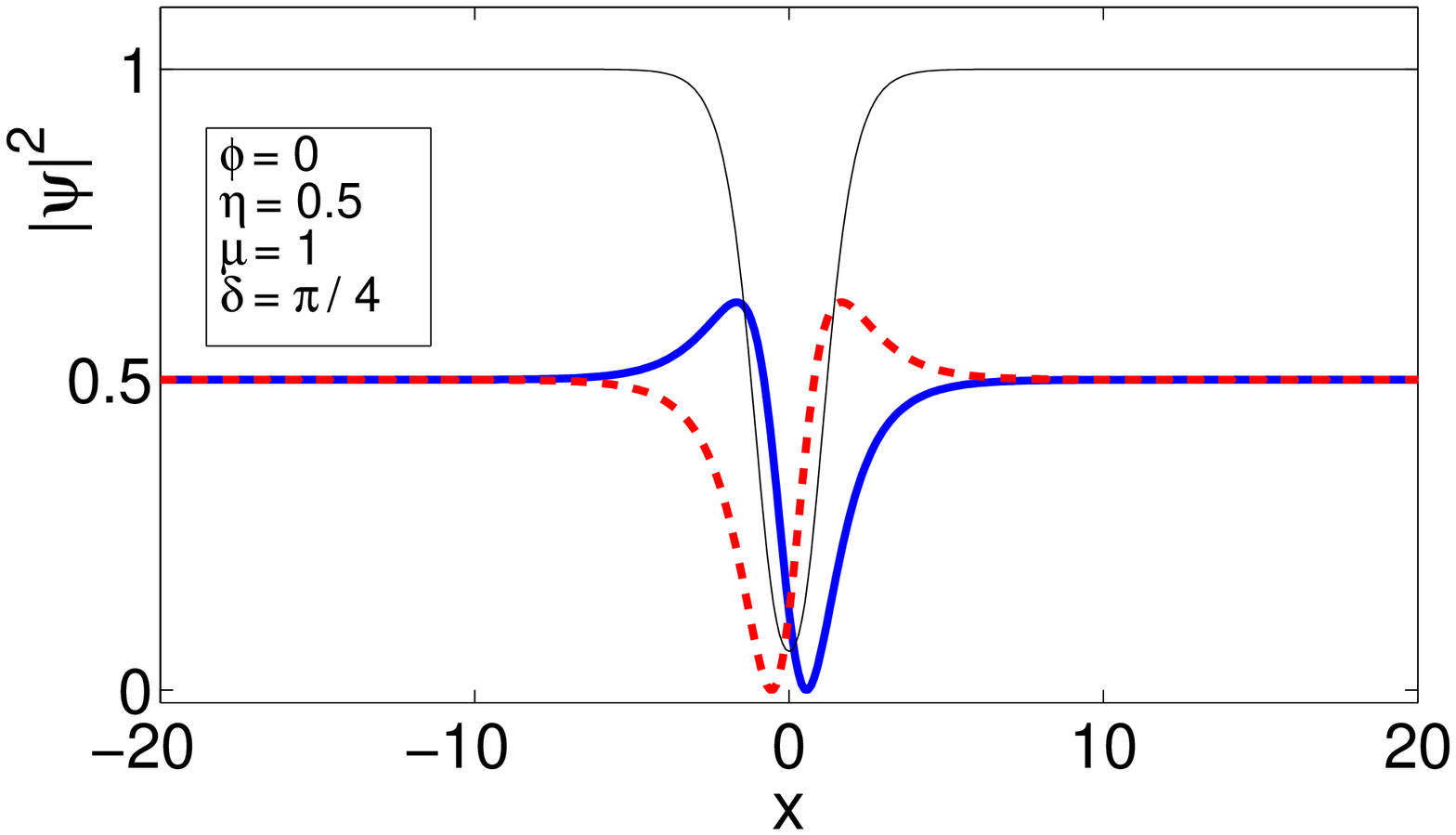}
\\
\includegraphics[scale=0.21]{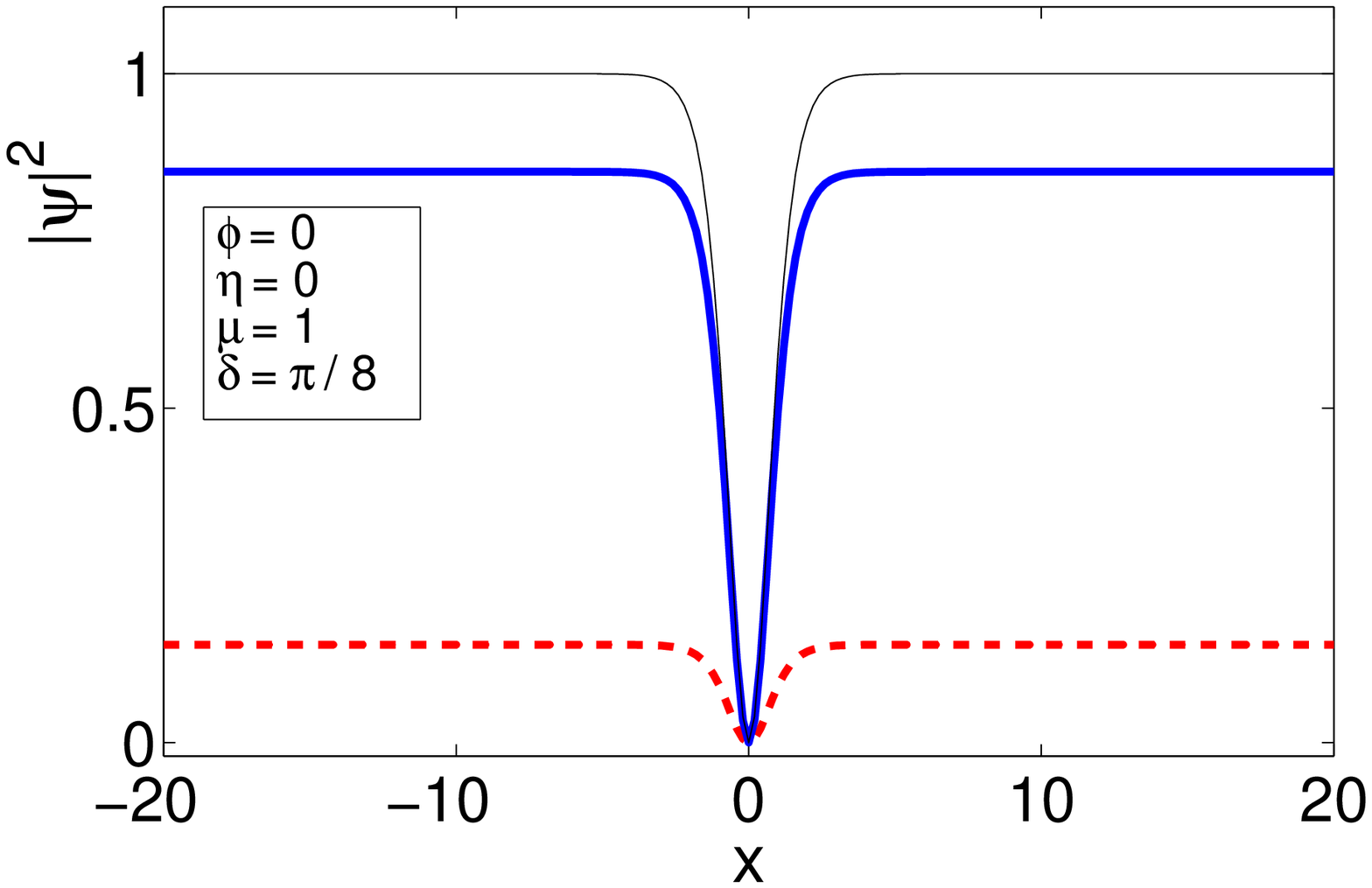}
			\includegraphics[scale=0.21]{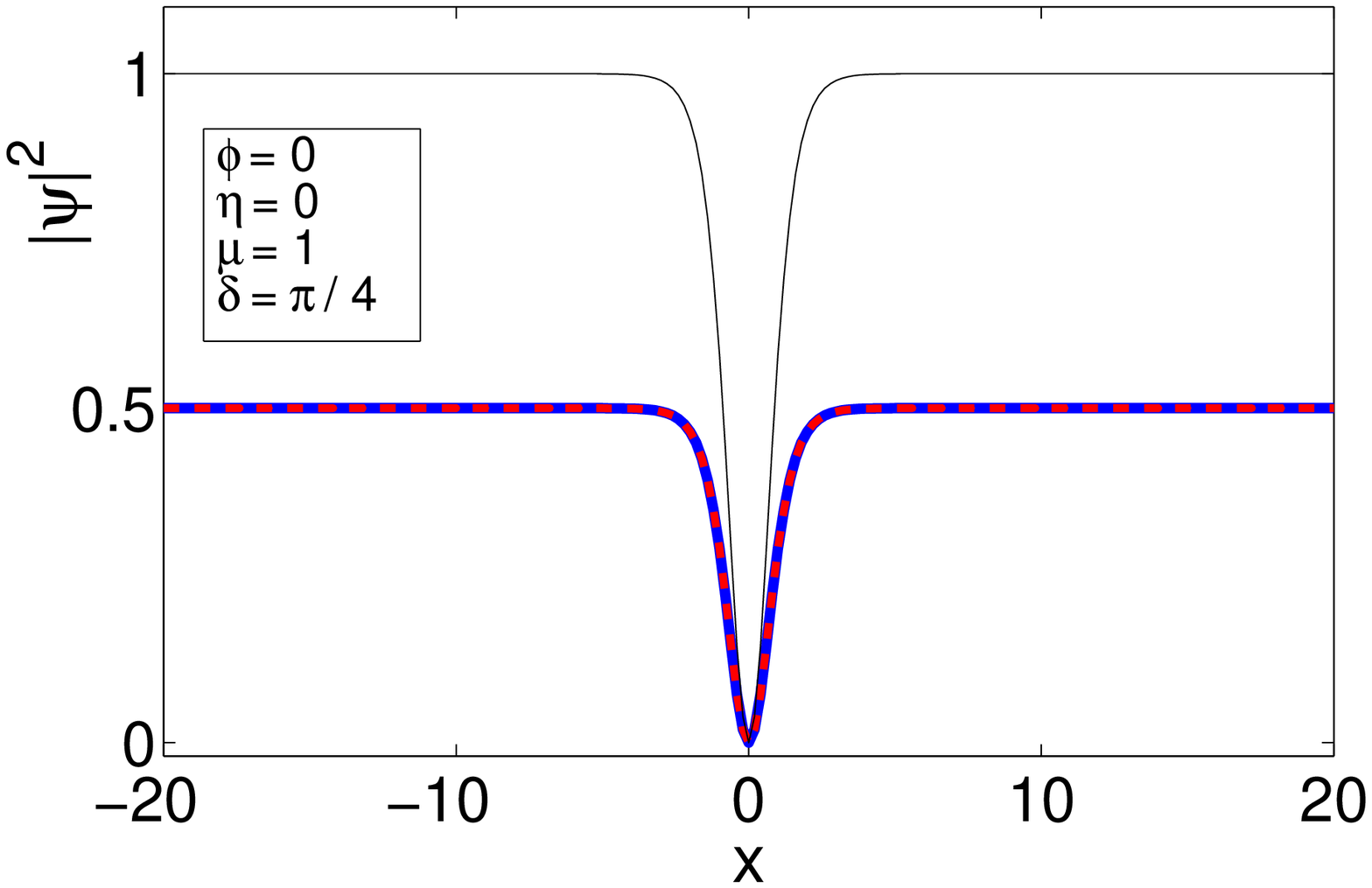}
		\end{tabular}
\caption{(Color Online)
Profiles of non-co-located
(for $\eta=0.5$; top panels), and
co-located ($\eta=0$; bottom panels) DD solitons.
For the left (right) panels,
$\delta =\pi/8$ ($\delta=\pi/4$), so density profiles
are asymmetric (symmetric).
Thick solid (dashed) line represents the first (second) component
and the thin solid line is
the total density (adapted from Ref.~\cite{siambook}).
}
		\label{revip_fig4}
	\end{center}
\end{figure}

A 
remarkable feature of the above Manakov system
is its invariance under SU$(2)$ rotations, i.e., under the action of
matrices of the form:
\[ U=\left( \begin{array}{ccc}
\alpha & -\beta^{*} \\
\beta & \alpha^{*} \end{array} \right), \]
with $|\alpha|^2 + |\beta|^2=1$, for complex $\alpha, \beta$ (stars denote complex conjugation).
For concreteness, limiting our considerations to the SO$(2)$ case~\cite{pe5},
we choose $\alpha=\cos(\delta)$ and $\beta=\sin(\delta)$. Then, the
respective (time-dependent) densities $n_i= |\psi_i|^2$
of the two components, upon rotation of a DB solution become:
\begin{eqnarray}
n_1 
&=& \mu \cos^2(\delta) -(\mu \cos^2(\delta) \cos^2 \phi
- \eta^2 \sin^2(\delta)){\sech}^2 \xi -\sqrt{\mu} \eta \sin(2\delta)\quad
\nonumber \\[0.25ex]
&\times& \left\{\sin\phi \sin[kx+\theta(t)] + \cos\phi \cos[kx+\theta(t)]\tanh \xi \right\}
{\sech}\xi,
\label{den1}
\\[2.0ex]
n_2 
&=& \mu \sin^2(\delta) -(\mu \sin^2(\delta) \cos^2 \phi
- \eta^2 \cos^2(\delta)){\sech}^2 \xi +\sqrt{\mu} \eta \sin(2\delta)\quad
\nonumber \\[0.25ex]
&\times& \left\{\sin\phi \sin[kx+\theta(t)] + \cos\phi \cos[kx+\theta(t)]\tanh \xi \right\}
{\sech}\xi.
\label{den2}
\end{eqnarray}
There are three key features to discern within these
complicated expressions.
This DD soliton family contains the ``standard''
co-located DD solitons in the
$\eta=0$ limit (i.e., in the absence of the bright component).
Second, the DD solitons are generally {\it asymmetric},
unless $\delta=\pi/4$. 
Third, the time-dependence
of phase 
$\theta(t)$ results in 
{\it time-dependent density profiles}, unless $\eta=0$; thus, DD solitons 
may feature
time-{\it dependent} densities.
For $\eta \neq 0$, the time-dependence is harmonic, 
with
an oscillation frequency $\omega_0$ such that:
\begin{eqnarray}
\frac{1}{2}k^2 < \omega_0 = \frac{1}{2}(k^2+D^2)
= \frac{1}{2}(\mu-\eta^2\sec^2 \phi) < \frac{1}{2} \mu,
\label{om}
\end{eqnarray}
These important 
features highlighted above
are illustrated in Fig.~\ref{revip_fig4}.

\subsection{Dark-bright and dark-dark solitons in the trap.}

We now turn to the case where a parabolic trapping potential is present.
In this setting, 
the experimental findings depicted in
Fig.~\ref{revip_fig2}
(see also Ref.~\cite{buschanglin}) illustrate that a DB soliton 
oscillates in the trap, following the dynamics of a classical
harmonic oscillator.
Such a ``particle-like'' oscillation of the DB soliton can be treated
perturbatively~\cite{vaspra,f3} in the 
Thomas-Fermi (TF) limit~\cite{stringari}, where the wave
can be considered as a 
Newtonian particle 
inside an effective potential.

To elaborate on this, we
first recall that in the TF limit the ground state density of the first component
(assumed to carry the dark soliton) is $|\psi_{1{\rm TF}}|^2 = \max\{\mu-V(x),0\}$~\cite{stringari}. Then, denoting by $u_d$ and $u_b$ the wavefunctions of
the dark and bright soliton, we use $\psi_1 \rightarrow \psi_{1{\rm TF}} u_d$
and $|\psi_2|^2 \rightarrow \mu^{-1} |u_b|^2$, as well as 
$t \rightarrow \mu t$ and $x \rightarrow {\sqrt{\mu}}x$, and
derive from Eq.~(\ref{deq12}) the system~\cite{vaspra}:
\begin{equation}
\begin{array}{rcl}
i\partial_tu_d+\frac{1}{2}\partial^2_xu_d-\left(|u_d|^2+|u_b|^2-1\right)u_d &=& R_d,
\\[1.5ex]
i\partial_tu_b+\frac{1}{2}\partial^2_xu_b-\left(|u_b|^2+|u_d|^2-\tilde{\mu}\right)u_b &=& R_b,
\end{array}
\label{u_d_u_b}
\end{equation}
where we have assumed chemical potentials
for the dark and bright solitons $\mu_1=\mu_d=\mu$ and $\mu_2=\mu_b=\mu+\Delta$
(with the difference being $\Delta<0$), so that $\tilde{\mu}=1+\Delta/\mu$.
Finally, 
the perturbations $R_d$ and $R_b$ 
are given by:
\begin{eqnarray}
R_d &\equiv& (2\mu^2)^{-1}[2(1-|u_d|^2)V(x)u_d+V'(x)\partial_xu_d],
\label{R_d}
\\[0.5ex]
R_b &\equiv& \mu^{-2}[(1-|u_d|^2)V(x)u_b].
\label{R_b}
\end{eqnarray}
We now assume adiabatic evolution of the DB soliton in the presence of the
perturbations $R_{d,b}$, so that the soliton preserves its shape but its parameters
become unknown functions of time, namely:
\begin{eqnarray}
D^2(t)=\cos^2\phi(t)-\frac{1}{2} \chi D(t) \quad {\rm and}\quad
\dot{x}_0(t)=D(t)\tan\phi(t).
\label{s2}
\end{eqnarray}
Then, calculating the rate of change of the total energy of the system:
\begin{eqnarray}
E = \frac{1}{2}\int_{-\infty}^{+\infty}
|\partial_{x} u_d|^2+|\partial_{x} u_b|^2+(|u_d|^2+|u_b|^2-1)^2
- 2(\tilde{\mu}-1)|u_b|^2\, dx, 
\label{1D:energy}
\end{eqnarray}
we obtain:
\begin{eqnarray}
4\dot{D}D^2+
\chi D\sec^2\phi(\dot{D}+D\dot{\phi}\tan\phi) = 
\frac{\cos\phi}{\mu^2}
\left(\sin2\phi-
\chi D\sin\phi\right) V'(x_0).
\label{perturbed}
\end{eqnarray}
We thus derive the dynamical system of Eqs.~(\ref{s2}) and (\ref{perturbed}), which possesses
the fixed point: $x_{0,{\rm eq}}=0$,  $\phi_{\rm eq}=0$,
$D_{\rm eq} = \sqrt{1+\left(\frac{\chi}{4}\right)^2}-\frac{\chi}{4}$.
Linearizing around it, and assuming a parabolic trap of
strength $\Omega$, i.e., $V(x)=(1/2)\Omega^2x^2$, we find the equation of motion
for the DB soliton center:
\begin{equation}
\ddot{x}_0 = - \omega_{\rm osc}^2 x_0
\quad {\rm where} \quad
\omega_{\rm osc}^2 = \Omega^2 \left(\frac{1}{2}- \frac{\chi}{8\sqrt{1+(\chi/4)^2}} \right).
\label{eqmot}
\end{equation}

\begin{figure}
	\begin{center}
\includegraphics[scale=0.25]{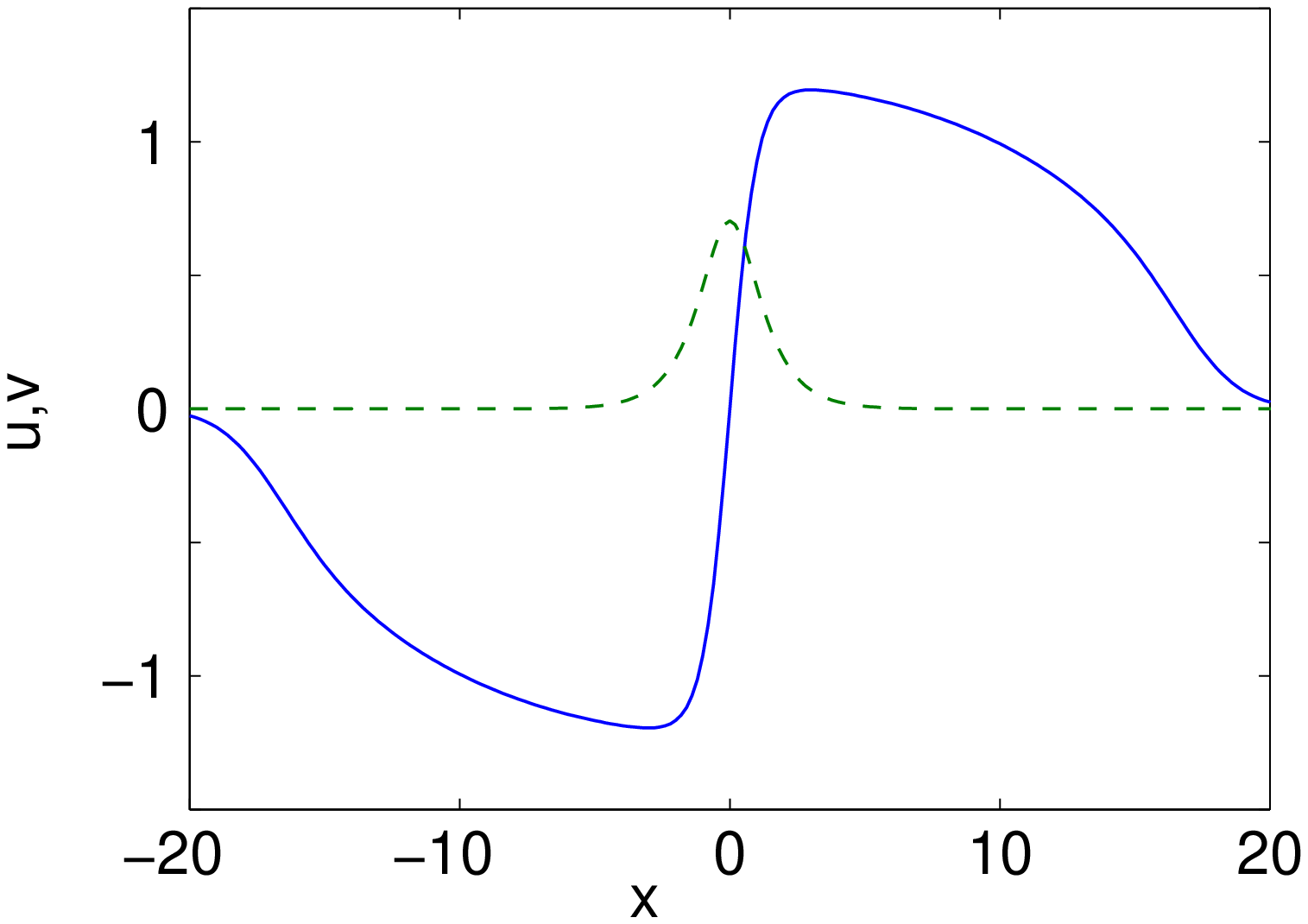}
\includegraphics[scale=0.25]{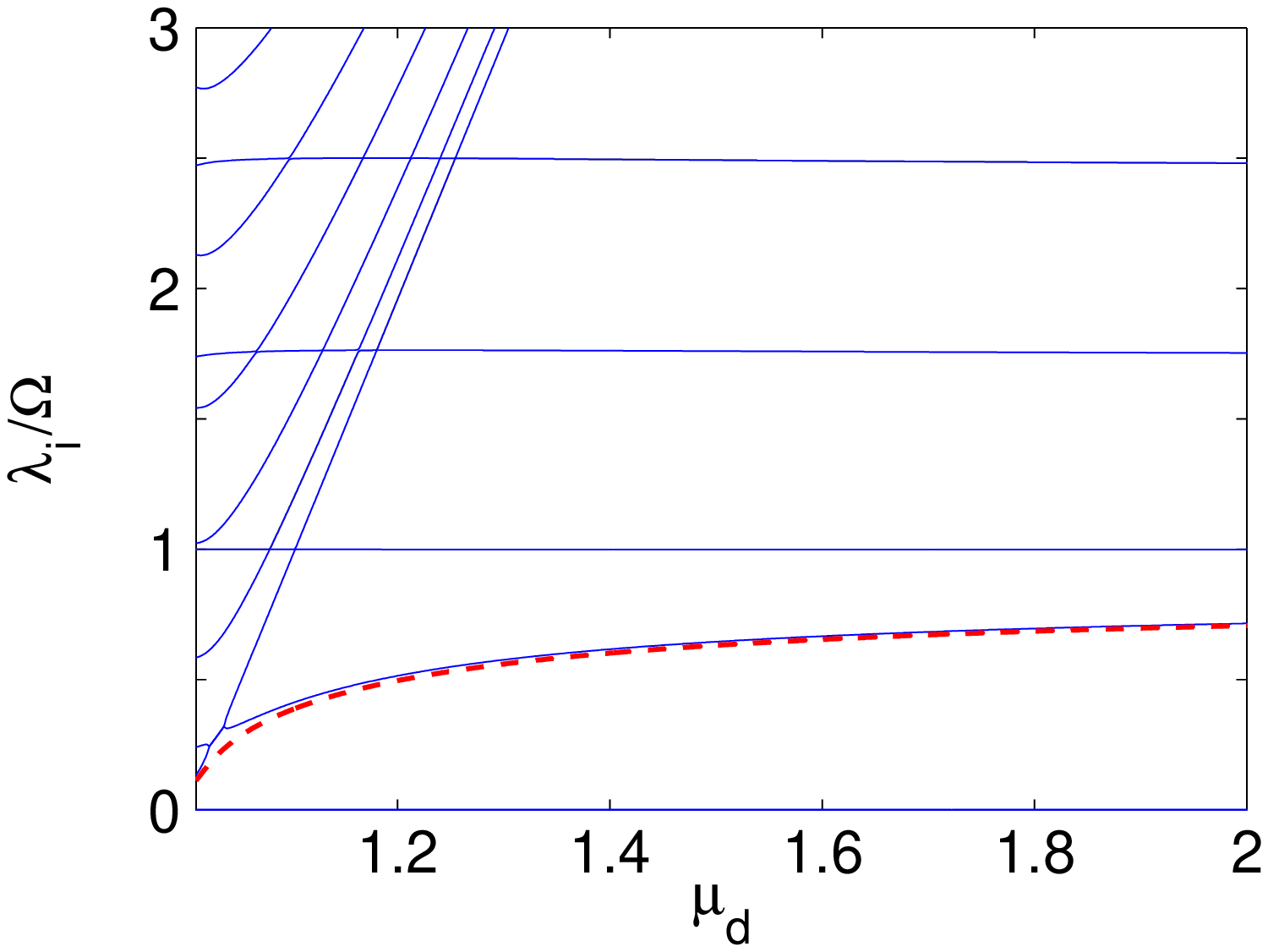}
\includegraphics[scale=0.25]{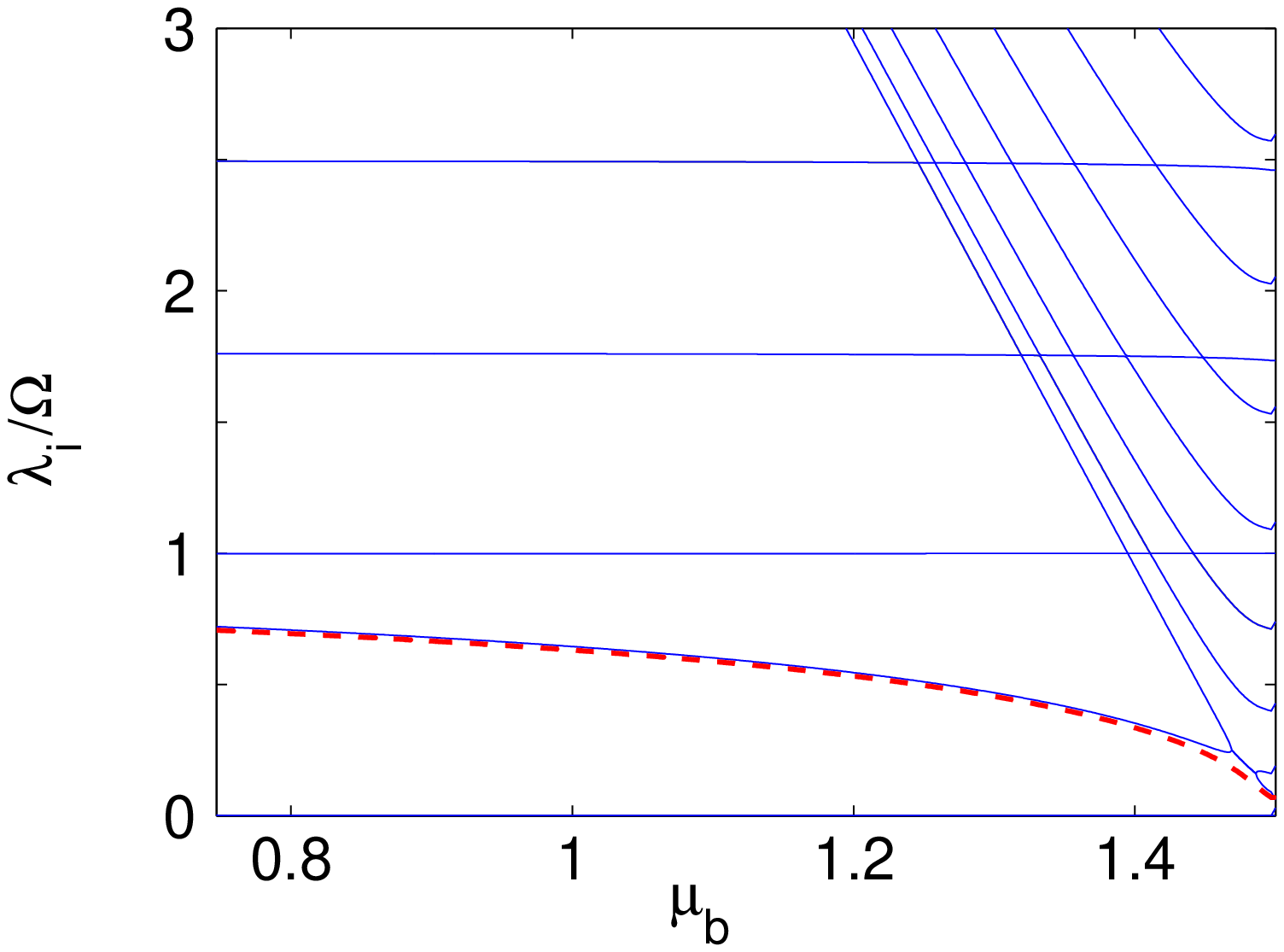}
		\caption{(Color Online) Left panel: a stationary
		DB soliton in a parabolic trap;
solid blue (dashed green) line depicts the dark (bright) component.
Middle and right panels: dependence of the
eigenvalues (normalized to the trap strength)
on the chemical potential of the dark and bright component, respectively, as
obtained numerically via the BdG analysis.
The lowest nonzero eigenfrequency
is in very good agreement with
the theoretical prediction (red dashed line) of Eq.~(\ref{perturbed})
(adapted from Ref.~\cite{pe3}). }
		\label{revip_fig5}
	\end{center}
\end{figure}

Note that the oscillation frequency can also be derived via a
Bogolyubov-de Gennes (BdG) analysis~\cite{stringari,siambook}
as follows.
Denoting by $u_{d,b}^{(0)}$ the dark and bright components of a stationary DB soliton
in the trap (cf.~left panel of Fig.~\ref{revip_fig5} for an example),
we use the ansatz
\begin{eqnarray}
u_d(x,t) &=& e^{-i \mu_d t} \left[ u_d^{(0)}(x) + \epsilon
\left(a(x) e^{\lambda t} + b^{\star}(x) e^{\lambda^{\star} t} \right)\right],
\label{darkp}
\\
u_b(x,t) &=& e^{-i \mu_b t} \left[ u_b^{(0)}(x) + \epsilon
\left(c(x) e^{\lambda t} + d^{\star}(x) e^{\lambda^{\star} t} \right)\right],
\label{brightp}
\end{eqnarray}
where $(a,b,c,d)^T$ is the eigenvector of the perturbation
and $\lambda$ is its corresponding eigenvalue.
Substituting Eqs.~(\ref{darkp})-(\ref{brightp})
in the equations of motion, and linearizing in the small parameter $\epsilon$,
we derive an eigenvalue problem for $[\lambda,(a,b,c,d)^T]$.
The essence of the BdG analysis is that,
once this problem is solved and $\lambda$ are found,
if ${\rm Re}\{\lambda\}>0$
then the solution is unstable; else
it is spectrally stable.
The BdG analysis in our case reveals that the eigenvalues are
imaginary, attesting to the DB soliton stability. 

The middle and right panels of Fig.~\ref{revip_fig5} show the dependence of the
imaginary part of the lowest nonzero eigenvalue 
on the chemical potential of the dark and bright component, respectively.
It is observed that the 
lowest nonzero eigenfrequency,
associated with an internal mode (so-called {\it anomalous mode}~\cite{stringari,siambook})
of the wave describing its oscillation in the trap,
is in very good agreement with
the theoretical prediction of Eq.~(\ref{perturbed}).

The analytical approach 
presented above is rather general in its nature: 
this energy balance methodology can be used to study the
dynamics of coherent structures of different nature, and even different dimensionality,
in Hamiltonian models in the presence of 
perturbations.
On the other hand,
it is relevant to remark that the SU$(2)$ and SO$(2)$ transformations
are {\it unaffected} by the presence of the trap, and hence apply to the
case of DD solitons 
with time-dependent density (so-called ``beating DD solitons''~\cite{pe5})
as they do for the DB ones.
Hence, the beating DD solitons 
have the same oscillation frequency [cf.~Eq.~(\ref{eqmot})] as the DB ones~\cite{pe5}.
Lastly, we note
that the above BdG results for the
1D setting, 
indicate that the DB soliton
is generally robust/spectrally stable;
however, in the full 3D setting
DB solitons turn out to be more prone to instabilities~\cite{pe2}.


\subsection{Multiple vector solitons and solitonic gluons 
}

The next experimental aspect that we address 
is that of the emergence
of bound states of multiple DB solitons. 
Such {\it solitonic gluons} (or ``soliton molecules'') were observed first
in optics~\cite{seg2} and also, 
more recently, in BEC experiments~\cite{pe3}. To provide an understanding for such states, we
employ the variational approach of Ref.~\cite{pe3}, and use the following ansatz for
two equal-amplitude DB solitons 
traveling in opposite directions:
\begin{eqnarray}
\!\!\!\!\!\!
u_d(x,t)
&\!\!=\!\!&\left(\cos\phi\tanh X_{-}+i\sin\phi \right)
\left(\cos\phi\tanh X_{+}-i\sin\phi\right),
\label{eq15}
\\[1.0ex]
\!\!\!\!\!\!
u_b(x,t)
&\!\!=\!\!& \eta\, {\sech} X_{-}\, e^{i\left[kx+\theta)+(\tilde{\mu}-1)t\right]}
+\eta\, {\sech} X_{+}\, e^{i\left[-kx+\theta)+(\tilde{\mu}-1)t\right]}\,
e^{i\Delta\theta}, \quad
\label{eq16}
\end{eqnarray}
where $X_{\pm} = D\left(x \pm x_0(t)\right)$, $2x_0$ is the relative distance between
the two waves, and $\Delta\theta$ is the relative phase between
the two bright components. Substituting the above ansatz into the energy
of the system [cf.~Eq.~(\ref{1D:energy})], and considering low-velocity and well-separated
($x_0\gg 1$) solitons, it is found that
the energy of the system assumes the form:
\begin{eqnarray}
E = 2E_1+ E_{\rm DD}+ E_{\rm BB} + 2E_{\rm DB},
\label{energyf}
\end{eqnarray}
i.e., the energy consists of twice the energy $E_1$ of one 
DB soliton,
the interaction energies $E_{DD}$ and $E_{BB}$ between the two dark and the two bright solitons,
and twice the cross interaction energy $E_{DB}$ of the dark component of one soliton
with the bright of the other (cf.~Ref.~\cite{pe3} for 
expressions of these energies).
We can then find the evolution of the soliton parameters from the energy conservation, $dE/dt=0$,
as in the previous case of the single DB soliton in the trap. This way,
for low-velocity, almost black solitons, 
energy conservation leads to the following equation for the soliton center:
\begin{eqnarray}
\ddot{x}_0 = F_{\rm int} \equiv F_{\rm DD}+F_{\rm BB}+2F_{\rm DB},
\label{Fint}
\end{eqnarray}
where we have used the same notation for the respective interaction forces.
To the leading order of approximation, it can be found that~\cite{pe3}:
\begin{equation}
F_{\rm BB} \propto {\rm e}^{-2 D_{\rm eq} x_0}\cos(\Delta\theta),
~~
F_{\rm DD}\propto {\rm e}^{-4 D_{\rm eq} x_0},
~~
F_{\rm DB} \propto {\rm e}^{-2 D_{\rm eq} x_0}\cos(\Delta\theta).
\end{equation}
The key feature 
here is that while the
DD interaction is always repulsive, the BB and DB ones
depend on the relative
phase $\Delta\theta$ between the bright components. In particular, if $\Delta\theta=0$
then the BB interaction is also repulsive, 
rendering impossible
a stationary ``molecular'' state. However, if $\Delta \theta=\pi$
(as 
found in Ref.~\cite{seg2}),
then the repulsive nature of the DD interaction dominates at short
distances, while the attractive nature of the BB interaction at long
ones, yielding the potential for a ``bound state'', the solitonic
gluon.

\begin{figure}
	\begin{center}
\includegraphics[height=2.6cm]{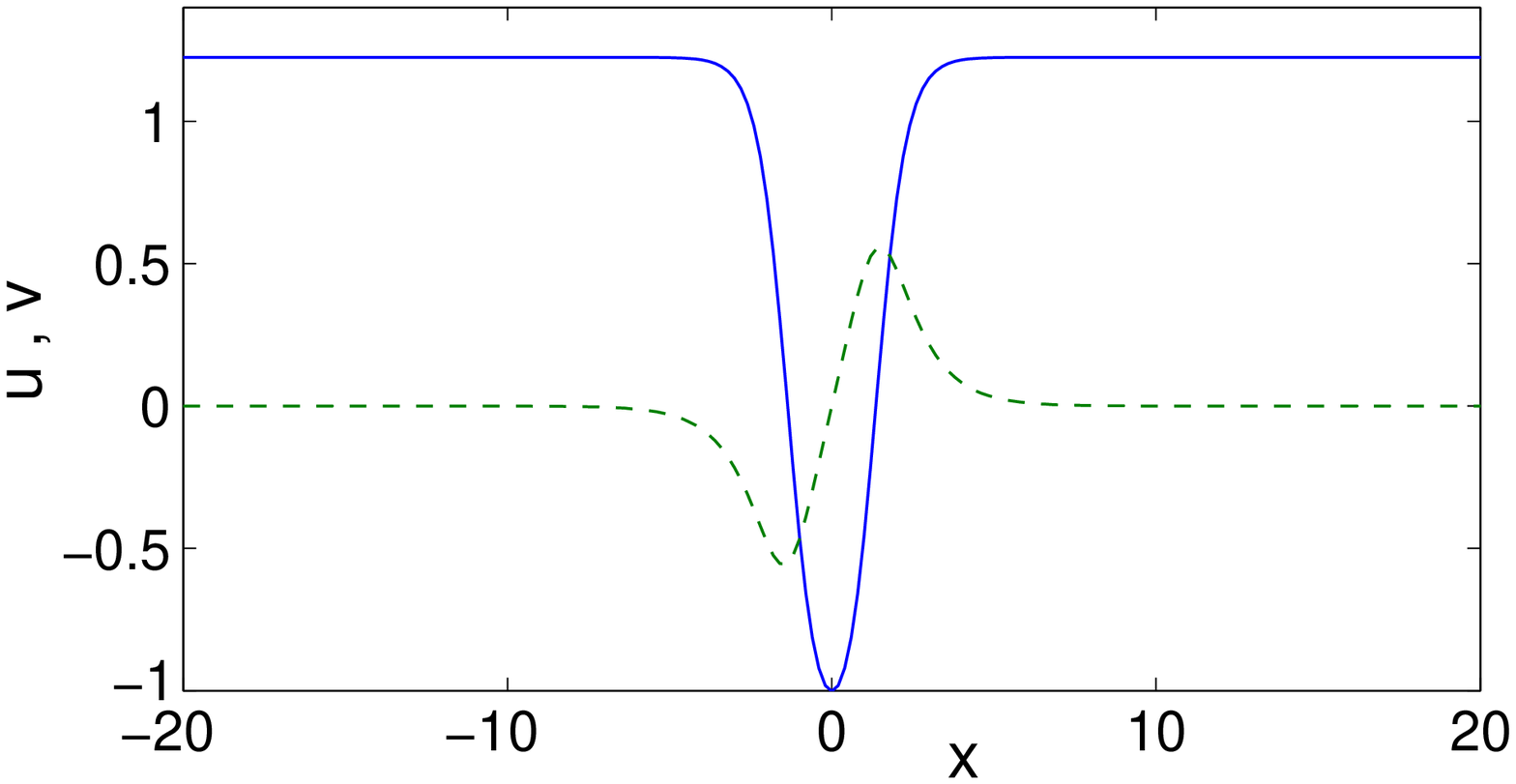}
\includegraphics[height=2.6cm]{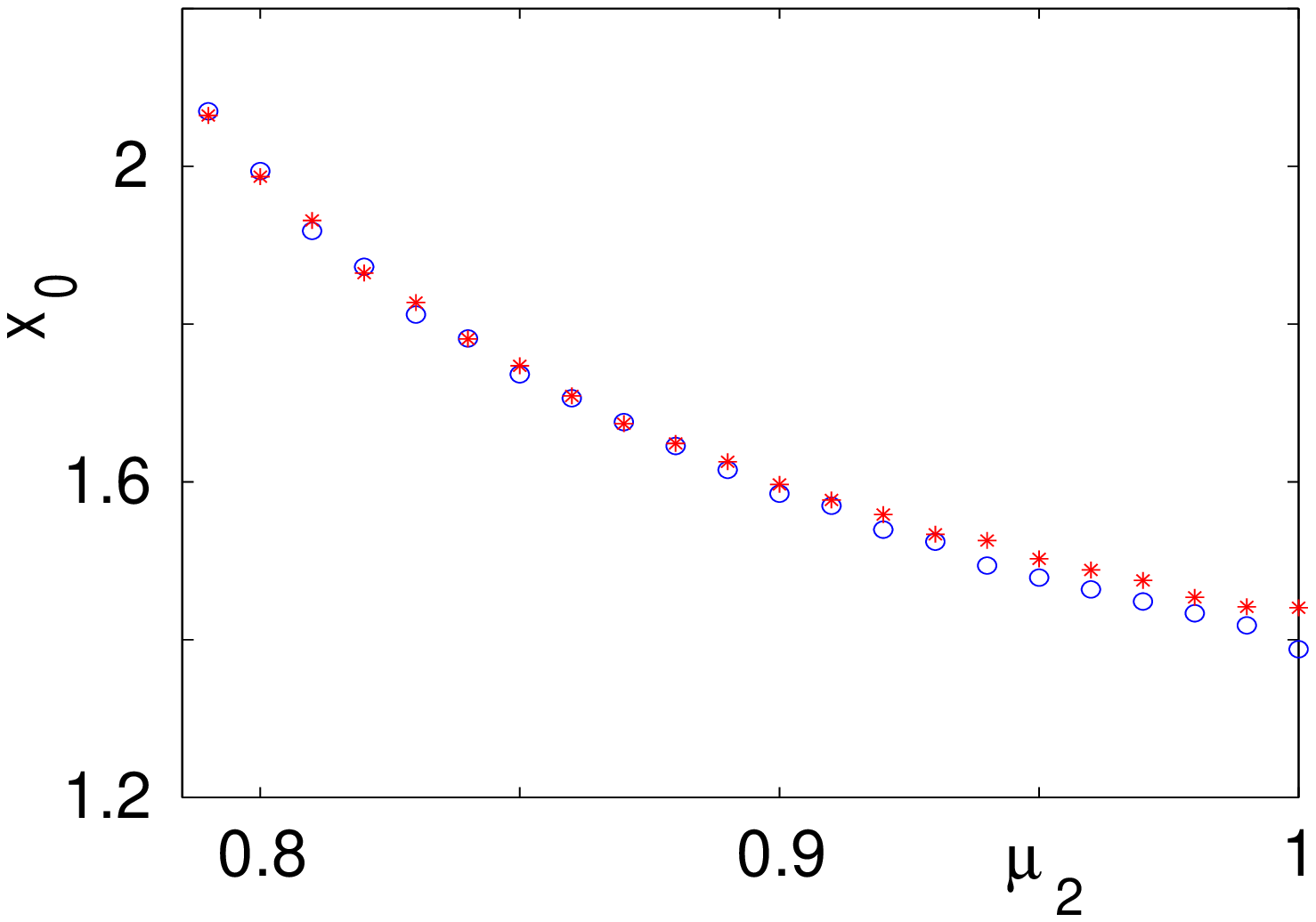}
\includegraphics[height=2.6cm]{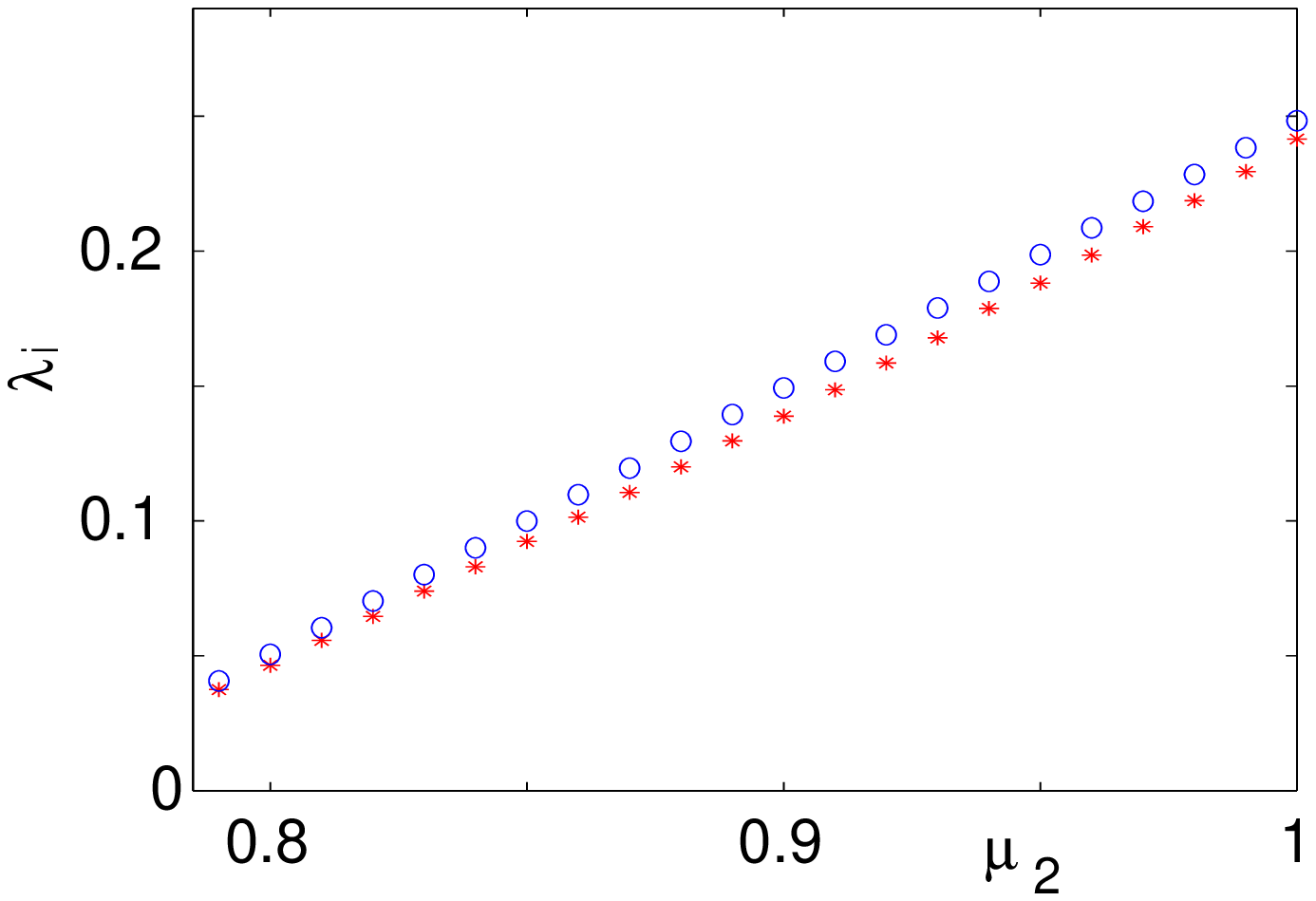}
		\caption{(Color Online) 
The dark (solid blue line) and bright (dashed green line) components of
a solitonic gluon in the homogeneous case (left panel).
The equilibrium distance between 
two DB solitons (middle panel)
and the out-of-phase oscillation frequency (right panel) as functions
of the bright component chemical potential;
red stars denote the theoretical prediction and blue
circles the numerical result (adapted from Ref.~\cite{pe3}). }
		\label{revip_fig6}
	\end{center}
\end{figure}

The above analysis allows for the identification of this state
numerically (even in the absence of a trap), 
and also provides the dependence of
the center of the solitonic gluon on,  e.g.,
the bright component chemical potential.
Equally importantly, linearization around
the equilibrium position $x_{\rm eq}$ (i.e., the distance between the constituent DB
solitons forming the stationary solitonic gluon)
suggests the existence of an internal mode of the two DB solitons,
involving their out-of-phase oscillation with a frequency 
$\omega_0^2= - \frac{\partial F_{\rm int}}{\partial x_0} \big|_{x_0 = x_{\rm eq}}$.
These predictions are illustrated in Fig.~\ref{revip_fig6}.

These molecular states can be generalized in the presence of a trap:
in this case, the equation of motion for the center $x_0$ of the solitonic gluon
involves not only the pairwise interaction force $F_{\rm int}$, but also
the restoring force of the trap $F_{\rm tr}$, inducing an in-trap oscillation with
a frequency $\omega_{\rm osc}$  [cf.~Eq.~(\ref{eqmot})].
Hence, the equation of motion for $x_0$ reads:
\begin{equation}
\ddot{x}_0 = F_{\rm tr} + F_{\rm int}.
\label{eqmot2}
\end{equation}
Thus, in this setting,
a molecular state exists even when the two DB solitons
are in-phase, because
their DD and BB (and DB) repulsion is counterbalanced by $F_{\rm tr}$.
An additional consequence is that, inside the trap, a pair of DB solitons 
will bear two, rather than one, internal modes.
The lowest one, 
pertains to their in-phase oscillation, 
characterized by the frequency $\omega_{\rm osc}$; 
the other internal mode 
corresponds to their out-of-phase motion, 
characterized by the frequency
$\omega_1^2 = \omega_0^2 + \omega_{\rm osc}^2$.

Motivated by the 
observation of
multi-DB-solitons
(cf.~right panel of Fig.~\ref{revip_fig2}), we may also consider
``lattices'' of such states. 
In particular, lattices of DB solitons, with
out- 
or in-phase bright neighbors respectively read~\cite{f3}:
\begin{eqnarray}
u_d &=& A_1 ~{\rm sn}(b x, k),  \quad
u_b = A_2 ~{\rm cn}(b x, k),
\label{latti1}
\\
u_d &=& A_1 ~{\rm sn}(b x, k),  \quad
u_b = A_2 ~{\rm dn}(b x, k),
\label{latti2}
\end{eqnarray}
where suitable (explicit) conditions connect 
amplitudes,
$A_1,~A_2$, and width as well as
inter-soliton separation parameters $b$ and $k$.
Such solutions exist
for general nonlinearity coefficients $g_{ij}$, and can be
numerically found
even well beyond
the regime of validity of
the analytical 
solutions~\cite{f3}.

Furthermore, in Ref.~\cite{wen}, it was found that the lattice states
can still be described via the above particle model
characterizing the pairwise interactions between the nearest-neighbor DB solitons. 
Moreover, a notion of ``kinetic temperature'' was introduced,
depending on the solitons' initial kinetic energy. 
In that light, a (gradual) transition of the dynamics of a large
number of solitons
could be identified as follows. When
the kinetic energy (and hence the kinetic temperature) of the system
was low, the array of the DB
solitons behaved as a crystal. As the
kinetic temperature was increased, the DBs progressively demonstrated a more
gaseous behavior with a large number of collision events.
Intriguingly, these also included an exchange of mass between
the bright components, rendering the particle description less
accurate in this limit. This is a topic worthwhile of further exploration.

\section{Variations on the theme}

We now turn to a number of case examples in which variants of the standard
homogeneous or trapped DB
solitons are of relevance/interest.

\subsection{The case of unequal masses}

\begin{figure}
	\begin{center}
		\begin{tabular}{ccc}
			\includegraphics[scale=0.21]{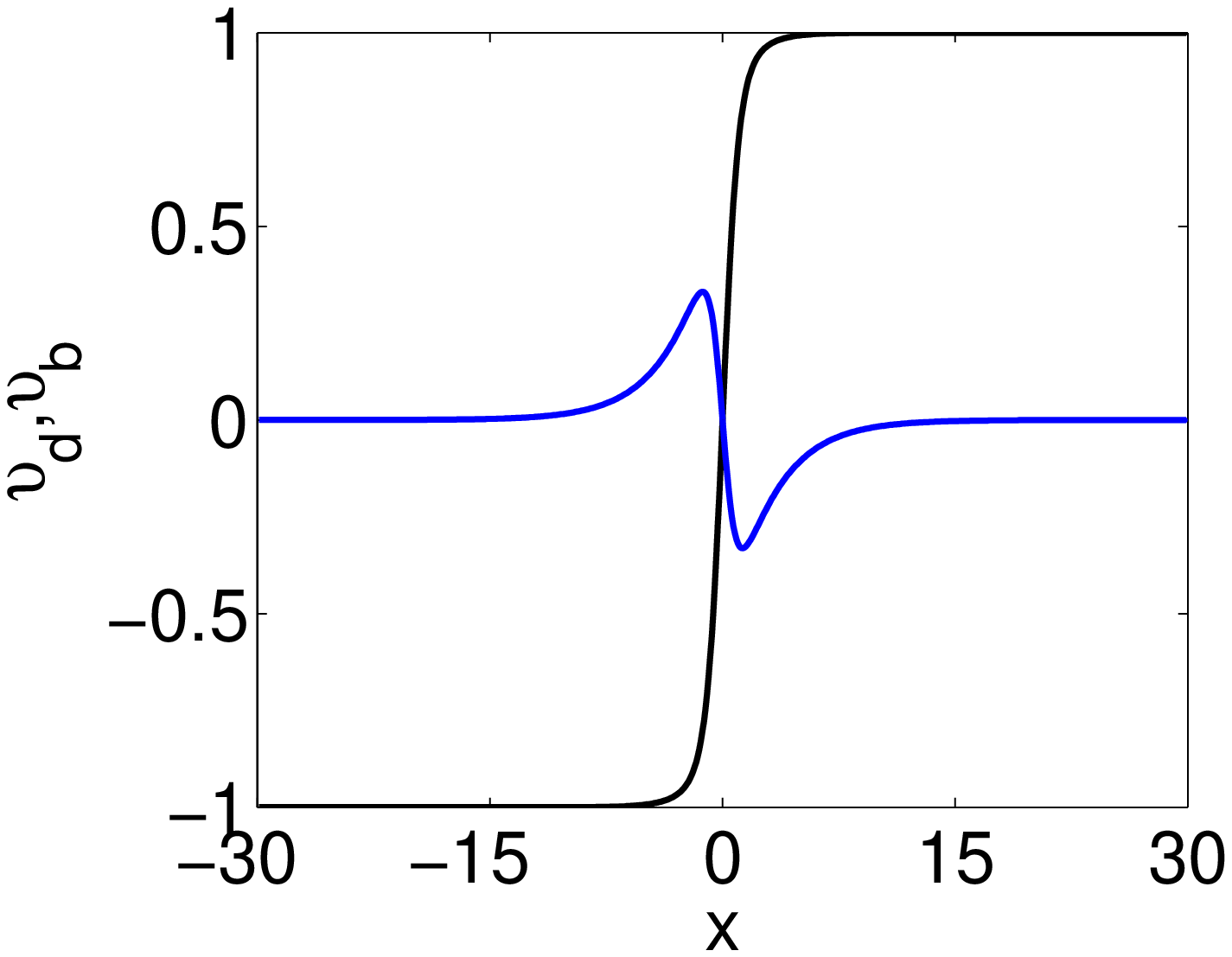}
			\includegraphics[scale=0.21]{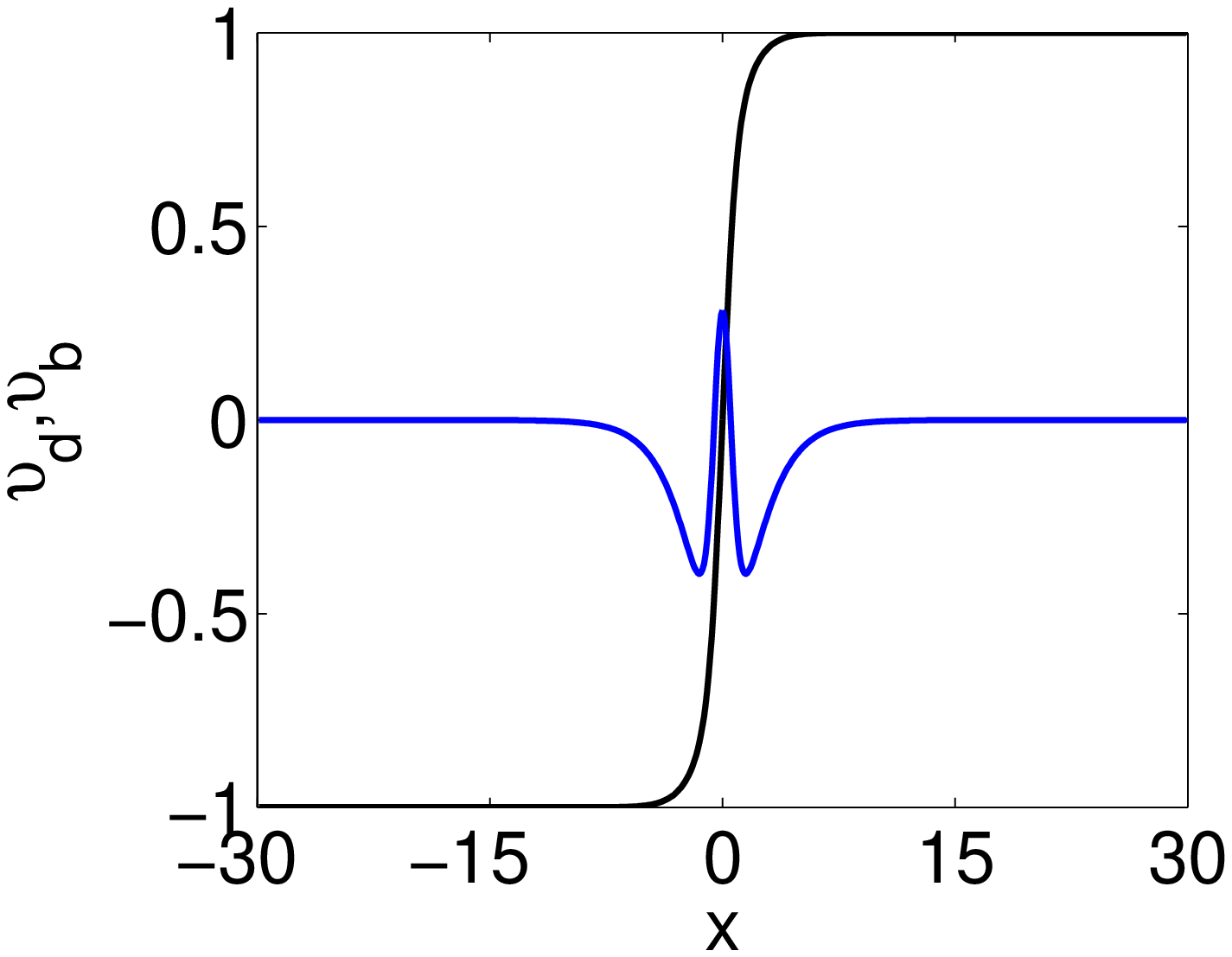}
\includegraphics[scale=0.21]{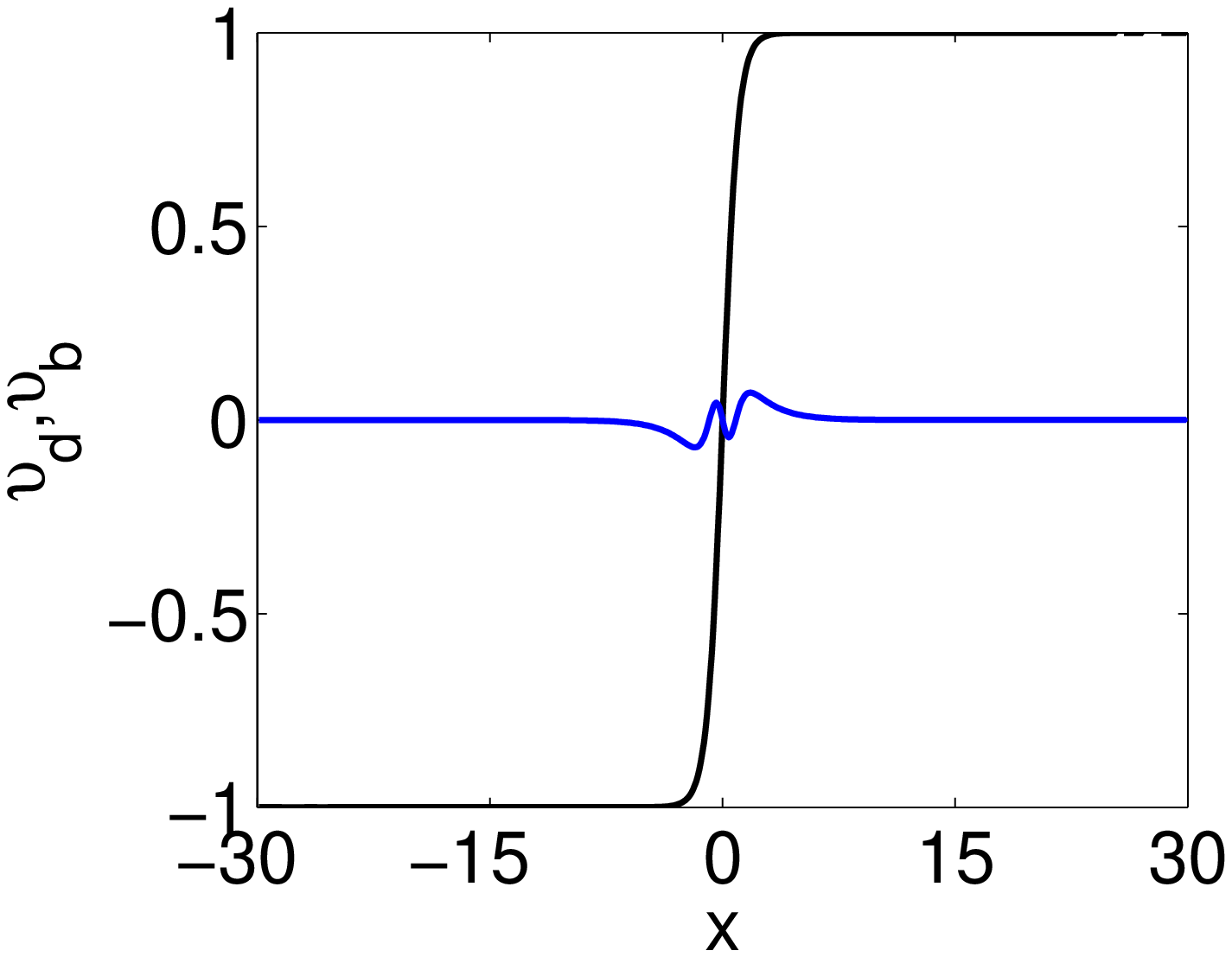}
		\end{tabular}
		\caption{(Color Online)
Excited states of bright solitons trapped by the dark soliton for $D<1$.
Left, middle and right panels show, respectively, the first,
the second (for $D<1/3$), and the third (for $D<1/6$) excited state.
The dark and bright component
profiles are 
depicted by black and blue lines, respectively (adapted from Ref.~\cite{s2}).
 }
		\label{revip_fig7}
	\end{center}
\end{figure}

Motivated by studies of spin-orbit coupled BECs~\cite{spinorbit1,spinorbit2},
it was recently shown~\cite{vaszb} that pertinent GPEs can be reduced to an
effective Manakov-type system but with
``unequal masses'' (i.e., unequal dispersion coefficients)
among the two components of Eq.~(\ref{deq12}).
Assuming that $D$ is the mass ratio of dark and bright components,
and seeking stationary solutions of
the GPEs~(\ref{deq12}) of the form $\psi_{1,2}=\upsilon_{d,b}(x)\exp(-i\mu_{d,b}t)$,
we obtain the system: 
\begin{equation}
\begin{array}{rcl}
\mu _{d}\, \upsilon _{d} &=&-\frac{1}{2}\left( \upsilon _{d}\right) ^{\prime \prime
}+\left( \upsilon _{d}^{2}+ \upsilon _{b}^{2}\right) \upsilon _{d}+V(x)\, \upsilon_{d},
\\[1.0ex]
\mu _{b}\, \upsilon_{b} &=&-\frac{D}{2}\left( \upsilon_{b}\right) ^{\prime \prime
}+\left( \upsilon_{d}^{2}+ \upsilon_{b}^{2}\right) \upsilon_{b}+V(x)\, \upsilon_{b}.
\end{array}
\label{-+}
\end{equation}
The existence of DB solitons 
can be explored upon realizing
that the dark soliton $u _{d}(x)=\sqrt{\mu _{d}}\tanh \left( \sqrt{\mu _{d}}x\right)$
[for $V(x)=0$] acts as an effective potential for the bright component~\cite{s2}.
This way, linearization of Eq.~(\ref{-+}) for a small bright component
gives rise to the eigenvalue problem:
\begin{equation}
\begin{array}{c}
\mathcal{L}\,\upsilon_{b}=
\lambda \,\upsilon_{b}
\quad {\rm with} \quad
 {\mathcal{L}=\frac{D}{2}}\frac{d^{2}}{dx^{2}}{+\mu _{d}\,
\mathrm{sech}^{2}\left( \sqrt{\mu _{d}}x\right) },
\end{array}
\label{quantum}
\end{equation}
where $\lambda=\mu_d-\mu_b$, while the Schr{\"o}dinger operator $\mathcal{L}$
corresponds to
the
P{\"o}schl-Teller potential, known from quantum mechanics~\cite{LL}.
Equation~(\ref{quantum}) supports 
bound states for 
integers $n$ satisfying
$D<D_{\mathrm{crit}}^{(n)}=\frac{2}{n\left( 1+n\right)}$.
Hence, 
a fundamental (bound) state
of the bright component, pertaining to the DB soliton, 
always exists. 
Furthermore, for $D<1$ (for $n=1$), it 
is also possible for the dark component
to trap a first excited state in the bright one 
(i.e., a two-bright-soliton anti-phase pair). For $D<1/3$, it
will be possible to trap a second excited state, and so on.
Examples of such states, which were identified in Ref.~\cite{s2}, are shown in Fig.~\ref{revip_fig7}.

\subsection{Dissipative dynamics under the action of thermal effects}

\begin{figure}
	\begin{center}
\includegraphics[height=4cm]{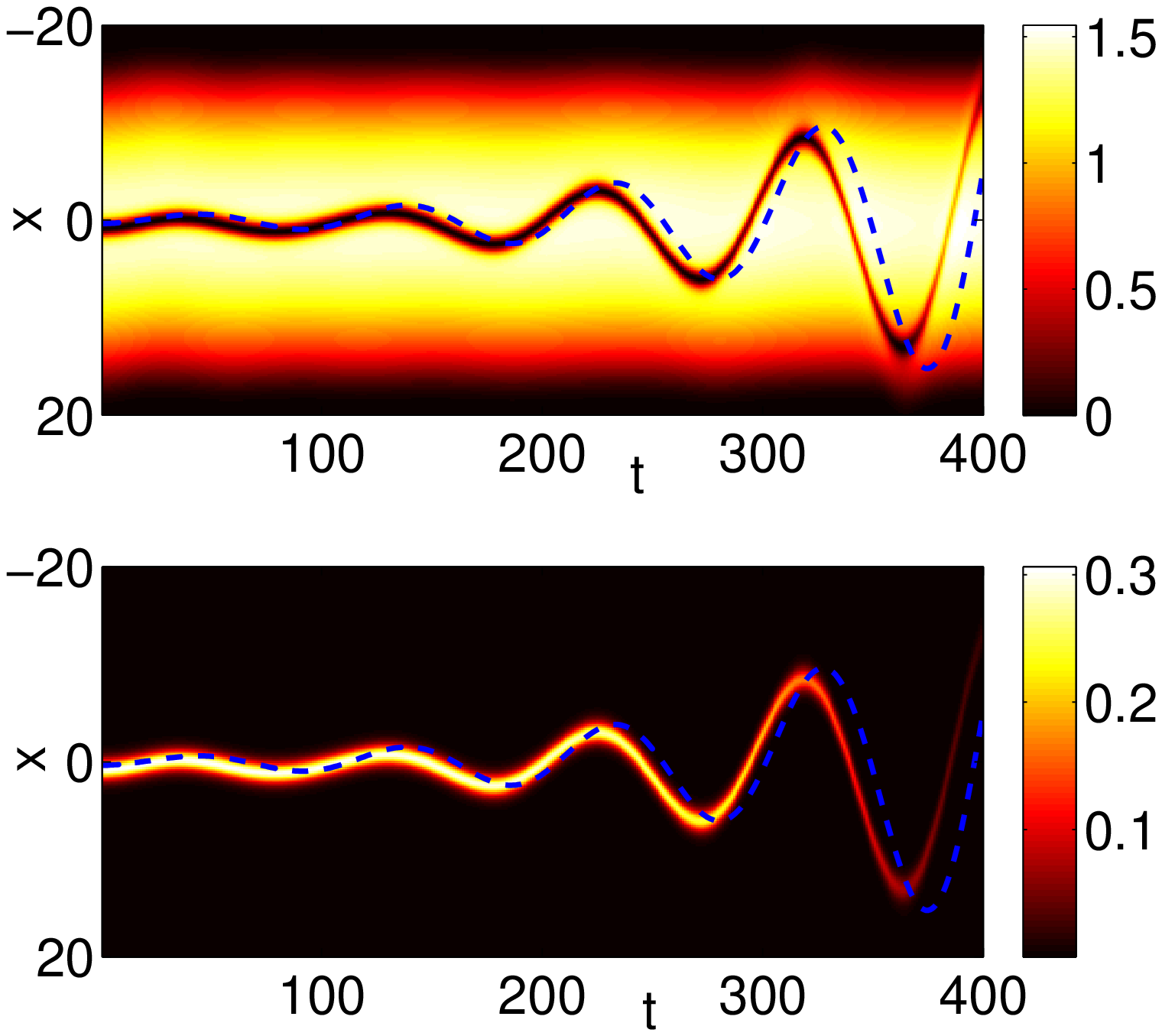}
\includegraphics[height=4cm]{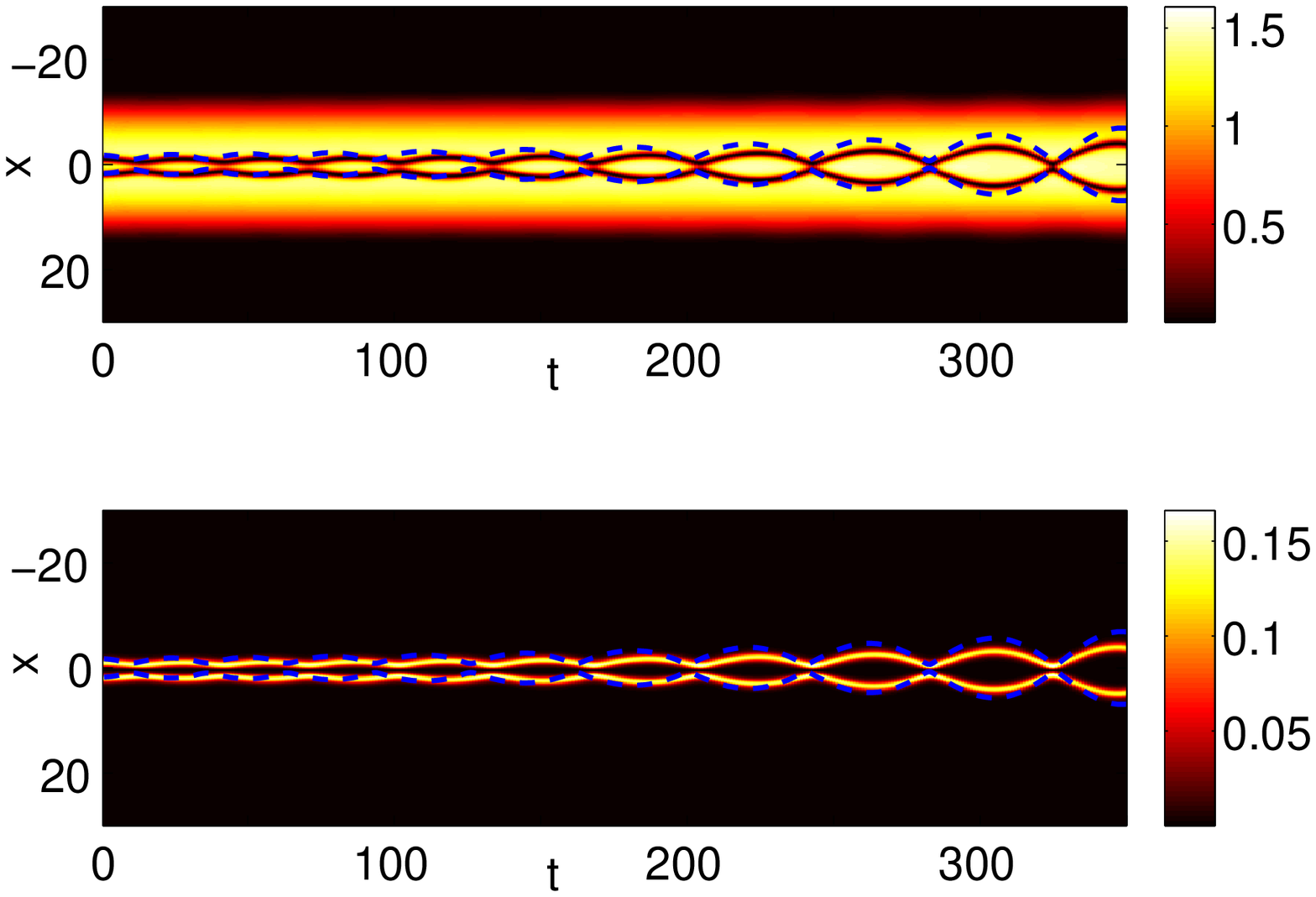}
		\caption{(Color Online)
Examples of the anti-damped dynamics of one (left panels) and two (right panels) DB
solitons; top (bottom) panels depict the dark (bright) component. Dashed (blue)
lines depict the analytical predictions (adapted from Ref.~\cite{vasnjp}).
 }
		\label{revip_fig8}
	\end{center}
\end{figure}

Dissipative effects, induced by the interaction of the BEC with the thermal cloud,
can be studied in the framework of the so-called {\it dissipative GPE} model~\cite{pitas}.
In the two-component setting, this model arises from Eq.~(\ref{deq12})
upon the substitution
$i \partial_t \psi_{1,2} \rightarrow (i- \gamma_{1,2})\partial_t \psi_{1,2}$~\cite{vasnjp}, where parameters $\gamma_{1,2}$ depend on
temperature~\cite{proukakis}.
The dissipative GPE system describes phenomenologically,
via the presence of losses, the transfer of atoms from the condensate
to the thermal cloud.
The dissipative dynamics of 
DB solitons in this setting can be studied
upon generalizing the methodology of Sec.~3.2. This way, 
it is possible to derive an equation for motion for the DB soliton center, of the form:
$\ddot{x}_0-a\,\dot{x}_0+\omega_{\rm osc}^2\,x_0 = 0$~\cite{vasnjp}, where
\begin{eqnarray}
a = \frac{2}{3}\mu\left(\gamma_1-
\frac{1}{8}\chi^2 \gamma_2\right) +\frac{1}{6}\mu
\left(\gamma_2-\gamma_1+
\frac{1}{8}\chi^2 \gamma_2\right)\frac{\chi}{\sqrt{1+(\chi/4)^2}}.
\label{aa}
\end{eqnarray}
The generic positivity of parameter $a$ 
suggests that the newly introduced term, $-a\dot{x}_0$, represents an effective {\it anti-damping}.
This term characterizes the interaction of the DB soliton 
with the thermal cloud. It results in the acceleration of the soliton
toward the velocity of sound, i.e., the dark component becomes continuously grayer and,
eventually, the wave transforms to the ground state of the BEC.
A similar situation,
in addition to their 
internal mode motion, is encountered in the case of multiple DB solitons. 
The accuracy of this effective particle picture in capturing
the anti-damped 
dynamics is shown
in Fig.~\ref{revip_fig8}.

It is important to note that, although this dynamics is similar to that of
dark solitons in single BECs~\cite{proukakis},
the analysis of Ref.~\cite{vasnjp} reveals that
the effect of the bright (``filling'') component is to partially
stabilize dark solitons against temperature-induced dissipation,
thus providing longer lifetimes.

\subsection{Dark-bright solitons in spinor condensates}

We now proceed with a case involving more than two components,
and study, more specifically, spinor BECs. The latter, have been realized
by employing optical trapping techniques, which allow for the confinement of
atoms regardless of their spin hyperfine state; thus, spinor BECs
formed by atoms with spin $F$, are described by a macroscopic wave function
with $2F+1$ components~\cite{stringari}. It is relevant
to highlight that spinor BECs
give rise to various phenomena that are not present
in single-component BECs, including formation of spin domains, spin textures,
topological states, and others~\cite{kawueda,stampueda}.

Here, we 
only consider a case example, namely a quasi-1D spinor $F=1$ BEC, described
by the following dimensionless mean-field model~\cite{ofymispin,bdspinor}:
\begin{eqnarray}
i\partial_{t}\psi_{\pm 1} &=&H_{0}\psi_{\pm 1}+\delta
\left[(|\psi_{\pm 1}|^{2}+|\psi_{0}|^{2}-|\psi_{\mp 1}|^{2})\psi_{\pm 1}
+\psi_{0}^{2}\psi_{\mp 1}^{\ast}\right] ,
\label{dvgp1} \\[1ex]
i\partial_{t}\psi_{0} &=&H_{0}\psi_{0}+\delta \left[(|\psi_{-1}|^{2}+|\psi_{+1}|^{2})\psi_{0}
+2 \psi_{-1}\psi_{0}^{\ast}\psi_{+1}\right].
\label{dvgp2}
\end{eqnarray}
Here, the components $\psi_{0,\pm1}$ correspond
to the three values of the vertical spin component $m_F=0,\pm1$, while
$H_{0}\equiv -(1/2)\partial_{x}^{2}+V(x)+|\psi_{-1}|^2+|\psi_{0}|^2+|\psi_{+1}|^2$,
and $\delta$ is the ratio of the strengths of the spin-dependent and
spin-independent interatomic interactions.
Note that $\delta$ is positive (negative) for {\it polar} ({\it ferromagnetic})
spinor BECs as, e.g., in the case of $^{23}$Na ($^{87}$Rb) atoms, where this
parameter takes the value $\delta =+3.14\times 10^{-2}$ ($\delta =-4.66\times 10^{-3}$).

As shown in Ref.~\cite{ofymispin}, for $\delta>0$
the background state (on which a dark soliton may be supported)
is modulationally stable; this suggests that DB soliton solutions
of Eqs.~(\ref{dvgp1})-(\ref{dvgp2}) may be possible. Indeed, in Ref.~\cite{bdspinor},
exploiting the smallness of $\delta$, a multiscale expansion method was used
[for $V(x)=0$] to show
that such states do exist, and assume the following form:
\begin{equation}
\begin{array}{rcl}
\psi_{\pm1}&=&\sqrt{\left( \mu /2\right) +\delta \rho(X,T)}
\exp\left[-i\mu t + (2i\delta/\mu) \int \rho(X) dX \right],
\\[1.0ex]
\psi_{0}&=& \delta^{3/4} q(X,T)\exp(-i\mu t),
\end{array}
\label{ps12}
\end{equation}
where $\mu$ is the chemical potential, $X=\sqrt{\delta}(x-\sqrt{\mu}t)$ and $T=\delta t$
are stretched variables,
while functions the $\rho(X,T)$ and $q(X,T)$ obey the following system:
\begin{equation}
\begin{array}{c}
\partial _{T}\rho =-\frac{1}{2}\sqrt{\mu}\partial_{X}\left( |q|^{2}\right), \quad
i\partial _{T}q+\frac{1}{2}\partial _{X}^{2}q-2\rho q=0.
\end{array}
\end{equation}
The above is the Yajima-Oikawa (YO) system, which was originally derived to
describe the interaction of Langmuir and sound waves in plasmas~\cite{yajima}.
This system is completely integrable, and possesses
soliton solutions of the form
$\rho \propto -{\rm sech}^2(k_s X-\omega_s T)$ and $q \propto {\rm sech}(k_s X-\omega_s T)$,
where $k_s,~\omega_s$ are constants. These expressions, when
substituted into Eq.~(\ref{ps12}), give rise to approximate
dark-dark-bright (for the $m_{F}=+1,-1,0$ spin components) solitons,
while a similar analysis can also lead to bright-bright-dark ones.
As shown in Ref.~\cite{bdspinor}, these small-amplitude structures
persist for large amplitudes and, in the presence of a parabolic trap,
they perform harmonic oscillations, in a way
reminiscent of the two-component case.

We note in passing that similar asymptotic reductions of other multi-component GPEs,
have been used to construct vector soliton solutions
described by integrable systems, such as the Yajima-Oikawa (already mentioned above)
and the Davey-Stewartson (DS) ones~\cite{aguero}, the Mel'nikov system~\cite{aguero,f1},
and a coupled Korteweg-de Vries (KdV) equations system~\cite{vectordark}.

\subsection{Vector solitons in higher-dimensions}

\begin{figure}[tbp]
	\begin{center}
			
\includegraphics[height=3.5cm]{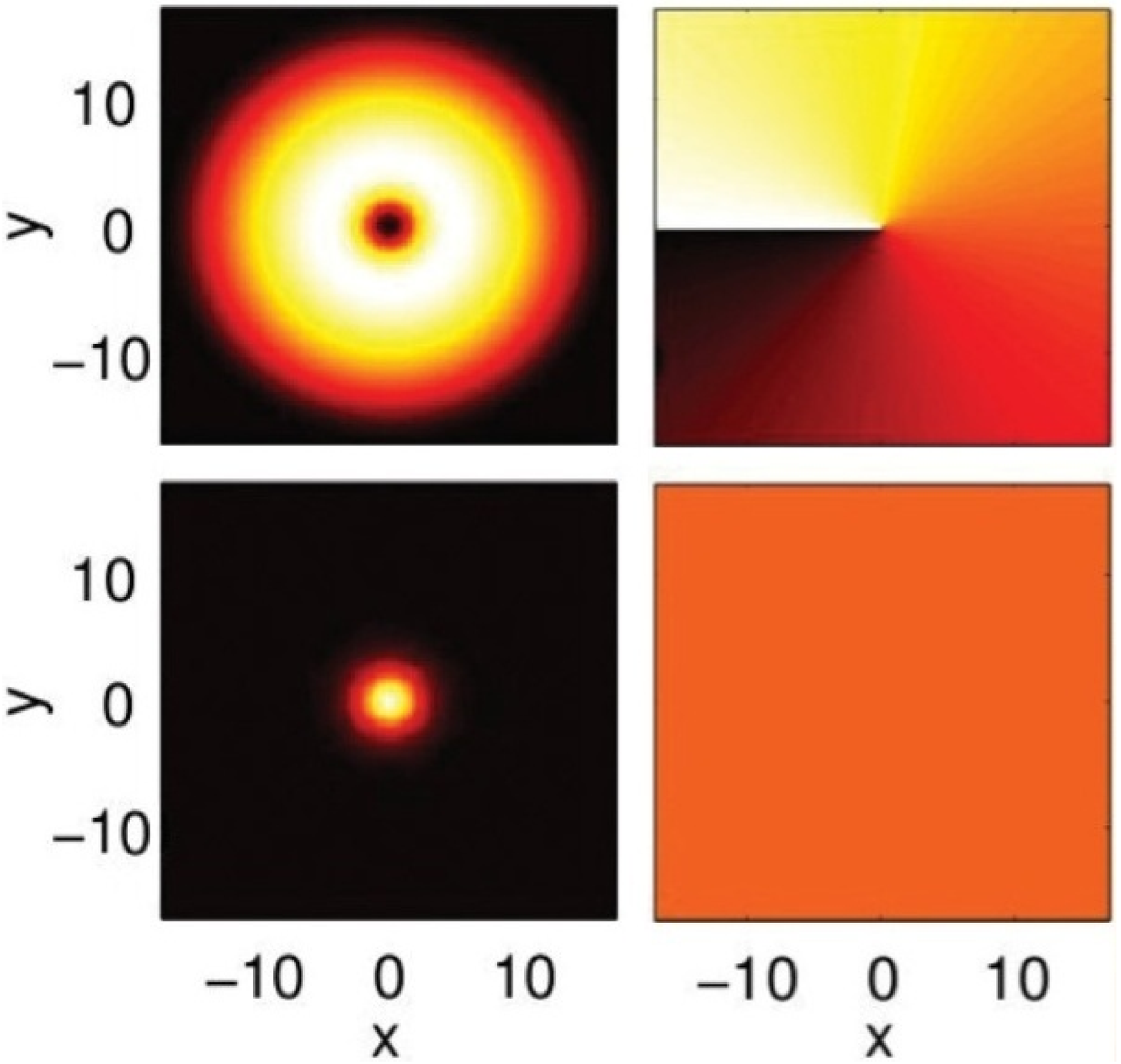}
\includegraphics[height=3.5cm]{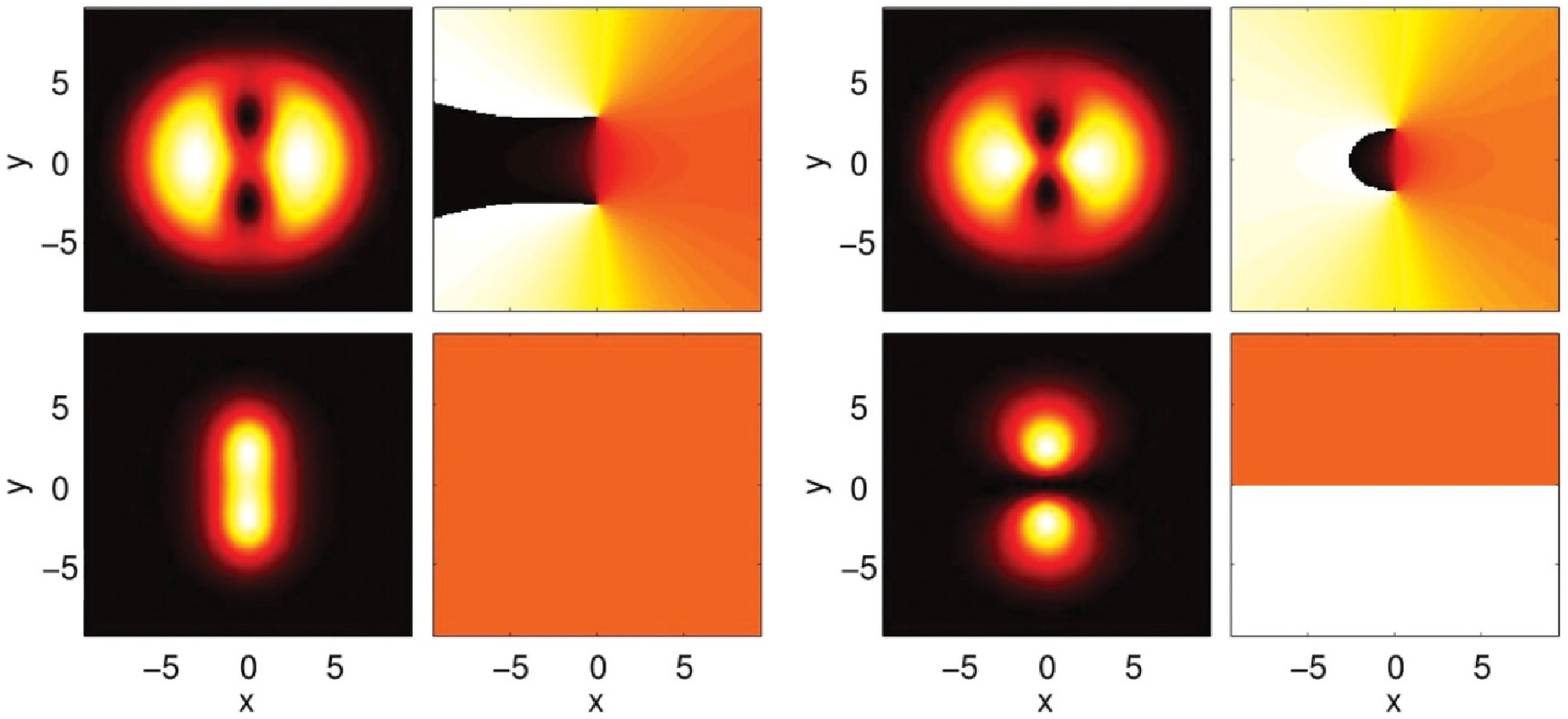}
		\caption{(Color Online)
The left panel quartet shows a single vortex-bright solitary wave, while
the middle and right quartets show, respectively, dipoles thereof,
with in-phase or out-of-phase bright components (adapted from Ref.~\cite{pola}).
Top (bottom) panels depict the vortex (bright) component, while
left (right) panels depict densities (phases).
 }
		\label{revip_fig9}
	\end{center}
\end{figure}

The
concept that one component
acts as a potential to trap the other,
is one that transcends dimensionality. Indeed, considering
the 2D variant of Eq.~(\ref{deq12}),
a vortex (which is a prototypical coherent structure
in 2D repulsive BECs~\cite{siambook}) in 
component $\psi_1$ can play the role of a potential
trapping a bright soliton in component $\psi_2$.
A case example of a vortex-bright soliton
is shown in Fig.~\ref{revip_fig9}. Note that these
structures bear different names in different communities, such as 
vortex-bright solitons~\cite{kodyprl,pola},
half-quantum vortices~\cite{tsubota} or baby Skyrmions~\cite{cooper}.
Various studies have been devoted to the stability~\cite{skryabin,kodyprl}
and dynamics~\cite{kodyprl,tsubota} of these structures.
Moreover, it was found that they feature 
intriguing interactions that decay as $1/r^3$~\cite{tsubota}.
They also allow much of the phenomenology discussed previously,
including their potential to form 
molecular states with out-of-phase and 
in-phase bright solitons in the presence of the trap  
(cf.~middle and right panels of Fig.~\ref{revip_fig9}, as well as the work of
Ref.~\cite{pola}).

Another quasi-2D 
state, 
is the ring DB soliton explored in Ref.~\cite{ringrds}. The matter-wave ring dark
soliton (RDS), introduced in Ref.~\cite{usrds}, 
is a 2D radial generalization of the dark soliton 
which, however, 
is generically unstable due to breakup into vortex-antivortex polygonal structures
(squares, hexagons, etc.). Nevertheless, in the two-component setting, the RDS in one component
can form an effective potential that supports a bright ring soliton structure
in the 
other component. As found in Ref.~\cite{ringrds}, although the presence of
the bright component 
weakens the instability of the RDS, it is not possible to eliminate it completely.
Nevertheless, the instability 
gives rise to states of interest in their
own right, such as vortex-bright polygons and DB soliton stripes.

\subsection{The double-well potential perspective}

Lastly, as regards the variants considered herein,
when two dark solitons
(or two vortices) trap
two bright ones, one can also view the relevant molecule
under a different prism, namely that of 
topological states in the first component forming an effective {\it double-well}
potential for the second one. This perspective was used
in the cases of DB solitons in 1D~\cite{vag},
and vortex-bright solitons in 2D~\cite{pola}. Importantly,
given their genuine topological nature,
vortices form, in a sense, 
a more ``robust'' double-well. Namely,
while this approach neglects
the back-action of the bright components on the dark ones
(i.e., this is a ``soft'' potential, rather than a hard,
externally imposed, one), 
this feedback mechanism is present and
is more significant in 1D than in 2D. Nevertheless,
in both settings, this approach enabled the observation of 
features associated with double-well potentials, such as
symmetry-breaking bifurcations, Josephson
oscillations, and so-called $\pi$-states emerging from
the bifurcations
(see Ref.~\cite{malomedbook} and references therein).

\section{Conclusions \& Outlook}

In this review, we examined
coherent structures arising
in coupled defocusing nonlinear Schr{\"o}dinger equations, especially so in
the vicinity of the so-called Manakov limit of equal self- and
cross-interactions. 
There (but also away from that
limit), a fundamental concept 
emerges, namely 
that a ``dark structure'' in one component (be it a dark soliton
in 1D or a vortex in 2D) 
acts as 
a potential well for the second component.
This enables the confinement therein of a bright soliton 
and the formation of ``dark-bright states''. These states also
persist in the presence of external (e.g., parabolic) potentials, 
wherein they 
oscillate as Newtonian particles. Molecular states involving multiple
dark-bright waves
may exist for
appropriate phase difference between the bright components in the homogeneous setting
(solitonic gluons), and in the presence of external traps.
Moreover, rotated versions of
such dark-bright solitons also emerge in experiments, in the
form of the so-called beating dark-dark solitons.

Additionally, 
variations on these themes were recognized and explored. 
The dark structure potential well was, for instance, recognized
as possibly trapping higher-excited states. Another possibility
concerned the
prototypical characterization of the thermal-induced dissipation, that leads
to anti-damping and eventual expulsion of 
dark-bright solitons 
from trapped atomic condensates.
The relevance of higher-component settings
bearing spinor analogues of the
considered states was also discussed.
Here, dark-dark-bright, or
bright-bright-dark states were found via a
multiscale asymptotic analysis. Another equally important and experimentally tractable
generalization arose in higher-dimensional settings, where
vortex-bright solitons, 
as well as dark-bright ring solitons
were presented.
Finally, 
molecular states bearing two (or more) dark entities were
also perceived as double (or, respectively, multi-) well potentials, thus
enabling
related phenomena.

We hope that it is clear 
that the above ideas and paradigms are powerful and broad beyond any one
of the particular examples used, and could lead to a wealth
of future possibilities and explorations, both in
theory and experiments.
As a small sample of
the questions that merit future consideration, we pose the
following. Is it possible to identify dark-bright solitonic
states (and their variants) in the context of the intensely studied
recently spin-orbit coupled 
BECs~? 
Can a quantitative
understanding of the scattering of vector solitons
off of Gaussian barriers (or wells)
be achieved~?
Can we trap higher-excited states in higher-dimensional settings
(radial and/or azimuthally dependent ones)~? Can an analogue
of dark-dark states be observed in higher-dimensions~? What
are the implications of these ideas in three-dimensional
settings~?
All these are important open questions,
a number of which are under active current consideration and, we
expect, will lead to numerous discoveries ahead...

\section*{Acknowledgments}

The invaluable contribution of all of our collaborators, co-authors and friends, 
including our past and present students and postdocs, 
is gratefully acknowledged. We are especially indebted to Ricardo Carretero-Gonz{\'a}lez 
for his extremely valuable input and help. The support of 
NSF-DMS-1312856 (PGK) is gratefully acknowledged.


\section*{References}

\bibliographystyle{elsarticle-num}


\end{document}